\documentclass[3p,twocolumn]{elsarticle}

\usepackage{circuitikz} 
\tikzset{>=latex} 
\usepackage{amsmath} 

\newcommand{\simm}{\raise.17ex\hbox{$\scriptstyle\sim$}}
\newcommand{\bmat}{\begin{bmatrix}}
\newcommand{\emat}{\end{bmatrix}}
\newcommand{\setof}[1]{\left \{ #1 \right \}}

\begin{document}
\begin{frontmatter}

\title{Field demonstration of predictive heating control for an all-electric house in a cold climate \\
{\it This is a preprint; full paper to appear in Applied Energy} }

\author[purdue]{Elias N. Pergantis\corref{correspondent}}
\author[purdue]{Priyadarshan}
\author[purdue]{Nadah Al Theeb}
\author[purdue]{Parveen Dhillon}
\author[emerson]{Jonathan P. Ore}
\author[purdue]{Davide Ziviani}
\author[purdue]{Eckhard A. Groll}
\author[purdue]{Kevin J. Kircher \corref{correspondent}}

\address[purdue]{Center for High Performance Buildings, Purdue University, 177 S Russell St, West Lafayette, IN 47907, USA}
\address[emerson]{Emerson Automation Solution, 200 Beta Drive, Pittsburgh, PA 15238, USA}
\cortext[correspondent]{Corresponding authors: \texttt{epergant@purdue.edu}, \texttt{kircher@purdue.edu}}

\begin{abstract}
Efficient electric heat pumps that replace fossil-fueled heating systems could significantly reduce greenhouse gas emissions. However, electric heat pumps can sharply increase electricity demand, causing high utility bills and stressing the power grid. Residential neighborhoods could see particularly high electricity demand during cold weather, when heat demand rises and heat pump efficiencies fall. This paper presents the development and field demonstration of a predictive control system for an air-to-air heat pump with backup electric resistance heat. The control system adjusts indoor temperature set-points based on weather forecasts, occupancy conditions, and data-driven models of the building and heating equipment. Field tests from January to March of 2023 in an occupied, all-electric, 208 m$^2$ detached single-family house in Indiana, USA, included outdoor temperatures as low as $-15$ $^\circ$C. On average over these tests, the control system reduced daily heating energy use by 19\% (95\% confidence interval: 13--24\%), energy used for backup heat by 38\%, and the frequency of using the highest stage (19 kW) of backup heat by 83\%. Concurrent surveys of residents showed that the control system maintained satisfactory thermal comfort. {\color{black} The control system could reduce the house's total annual heating costs by about \$300 (95\% confidence interval: 23--34\%).} These real-world results could strengthen the case for deploying predictive home heating control, bringing the technology one step closer to reducing emissions, utility bills, and power grid impacts at scale.
\end{abstract}


\end{frontmatter}

\section{Introduction}
\subsection{\color{black} Heat pumps and the power grid} 

Burning fossil fuels to heat buildings causes about five percent of global greenhouse gas emissions \cite{IEAbuildings}. Replacing fossil-fueled heating systems by heat pumps that run on low-carbon electricity could deeply reduce these emissions. However, heat pumps can sharply increase electricity demand during cold weather, when heat demand rises and heat pump efficiencies may fall. Peaks in electricity demand can stress power grid infrastructure, increasing the risk of power outages \cite{sharma2021major}. Expanding power grid capacity can mitigate the risk of power outages, but also incurs high costs \cite{horowitz2019distribution}, which utilities typically pass on to ratepayers \cite{EIA2023}. For these reasons, keeping electricity reliable and affordable while accommodating rapid heat pump adoption will likely require demand management strategies as an alternative or supplement to expanding power grid capacity \cite{zhou2021electrification}. This paper investigates one such strategy: equipping heat pumps with supervisory control systems that can reshape electricity demand while keeping building occupants comfortable. Supervisory heat pump control systems have other potential benefits, such as improving energy efficiency, reducing utility bills, reducing emissions of greenhouse gases and other pollutants, and providing power grid reliability services \cite{kircher2021heat}.

\subsection{\color{black} Heat pump control systems} 

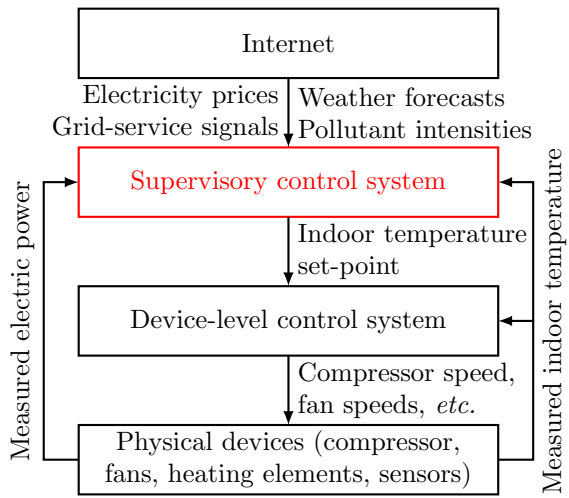
\begin{figure}
\centering
\begin{tikzpicture}[scale=0.92]
    \draw[thick] (0,0) rectangle (6,1);
    \node[align=center] at (3,0.5) {Physical devices (compressor, \\ fans, heating elements, sensors)};

    \draw[thick] (0,2) rectangle (6,3);
    \node[align=center] at (3,2.5) {Device-level control system};

    \draw[thick,red] (0,4) rectangle (6,5);
    \node[align=center,red] at (3,4.5) {Supervisory control system};

    \draw[thick] (0,6) rectangle (6,7);
    \node[align=center] at (3,6.5) {Internet};

    \draw[thick,->] (3,2) -- (3,1);
    \node[right,align=left] at (3,1.5) {Compressor speed, \\ fan speeds, {\it etc.}};

    \draw[thick,->] (3,4) -- (3,3);
    \node[right,align=left] at (3,3.5) {Indoor temperature \\  set-point};

    \draw[thick,->] (6,0.5) -- (6.5,0.5) -- (6.5,2.5) -- (6,2.5);
    \draw[thick,->] (6.5,2.5) -- (6.5,4.5) -- (6,4.5);
    \node[below,rotate=90] at (6.5,2.5) {Measured indoor temperature};
    \draw[thick,->] (0,0.5) -- (-0.5,0.5) -- (-0.5,4.5) -- (0,4.5);
    \node[above,rotate=90] at (-0.5,2.5) {Measured electric power};

    \draw[thick,->] (3,6) -- (3,5);
    \node[right,align=left] at (3,5.5) {Weather forecasts \\ Pollutant intensities};
    \node[left,align=right] at (3,5.5) {Electricity prices \\ Grid-service signals};
    
\end{tikzpicture}
\caption{A heat pump control architecture. This paper develops the supervisory control system (red block).}
\label{architectureFig}
\end{figure}

This paper considers the heat pump control architecture shown in Fig. \ref{architectureFig}. In this architecture, the device-level control system adjusts operating parameters, such as compressor and fan speeds, to keep the measured indoor temperature near a set-point. Device-level control systems typically use Proportional-Integral-Derivative loops with additional logic that protects equipment. This paper develops a supervisory control system that adjusts indoor temperature set-points. While sophisticated device-level control systems optimize thermal comfort and energy efficiency, they usually do not consider peak demand, dynamic prices, emissions, or grid-service revenues. Supervisory control systems can work in concert with device-level control systems to balance these higher-level objectives.

\subsection{\color{black} Supervisory HVAC control methods} 

Researchers have evaluated a variety of supervisory control methods for electric heating, ventilation and air conditioning (HVAC) equipment. Review papers such as \cite{killian2016ten, drgovna2020all, wang2020reinforcement, nagy2023ten} survey these research efforts, which date back at least to the 1980s \cite{stoecker1981reducing, braun1988methodologies, spratt1989dynamic}. Seminal early contributions such as \cite{braun1990reducing} focused on open-loop optimal control methods, which plan a trajectory of future set-points and implement them without feedback from real-time measurements or updated forecasts. Later, \cite{henze1997development} and subsequent work extended the open-loop optimal control approach to a framework known as model predictive control (MPC) \cite{kouvaritakis2016model}. Like open-loop optimal control, MPC plans a trajectory of future set-points. Unlike open-loop optimal control, MPC periodically re-plans as new measurements and forecasts arrive. In parallel with MPC research, a line of work tracing back to \cite{henze2003evaluation} investigated reinforcement learning control (RLC) for HVAC equipment. While MPC requires a mathematical model of a building's thermal dynamics, RLC can learn a model-free control policy directly from a reward or penalty signal.

\subsection{\color{black} Simulations vs. field demonstrations} 

Research on supervisory HVAC control can be categorized into commercial vs. residential applications, and into computer simulations vs. field demonstrations. To the authors' knowledge, research within the scope of this paper -- residential field demonstrations -- consists only of the 12 papers summarized in Table \ref{residentialExperimentTable}. For comparison, \cite{blum2022field} recently reviewed commercial MPC field demonstrations and identified 14 published examples. Similarly, \cite{nagy2023ten} recently found three published examples of commercial RLC field demonstrations. Collectively, field demonstrations comprise a small fraction of research on supervisory HVAC control systems. The review papers \cite{killian2016ten, drgovna2020all, wang2020reinforcement, nagy2023ten} emphasize the importance of field demonstrations as steps toward deploying supervisory control technology at scale. Field demonstrations can highlight practical implementation challenges that require further research efforts, evaluate real-world deployment costs and labor requirements, and help convince decision-makers in business and government that the technology works and is, or could soon become, economically viable.

\begin{table*}[t] 
\caption{Summary of field demonstrations of supervisory control systems for residential HVAC equipment.} 
\label{residentialExperimentTable} 
\centering
\small

\begin{tabular}{ p{0.7cm}  p{0.8cm}  p{2.0cm}  p{2.4cm}  p{2.4cm}  p{1.1cm}  p{1.5cm}  p{1.9cm} }
Study, Year & Control method & Building(s), Location & Equipment & Control action(s) & Duration & Objective & Objective improvement \\
\hline

\cite{pedersen2013central},  2013 & MPC & 4 houses in Denmark & Ground-to-water heat pumps & Compressor on/off state & 152 days & House energy cost & 9\% vs. measurement \\
\hline

\cite{dong2014real},  2014 & MPC & 1 house in the USA & Resistance radiant panels, air conditioner & Indoor temperature set-point, air flow & 10 days & HVAC energy & 18--30\% vs. simulation \\
\hline

\cite{afram2017supervisory},  2017 & MPC & 1 house in Canada & Ground-to-water heat pump & Indoor temperature set-point & 22 days & HVAC energy cost & 16--50\% vs. simulation \\
\hline

\cite{bunning2020experimental},  2020 & MPC & 1 bedroom in Switzerland & Radiant panels & Water flow & 6 days & HVAC energy & 25\% vs. measurement \\
\hline

\cite{finck2020optimal},  2020 & MPC & 1 house in the Netherlands & Air-to-water heat pump & Indoor temperature set-point & 3 days & HVAC energy cost & 12--15\% vs. measurement \\
\hline

\cite{kurte2020evaluating},  2020 & RLC & 1 house in the USA & Central air conditioner & On/off state and indoor temperature set-point & 5 days & HVAC energy cost & 7--17\% vs. simulation \\
\hline

\cite{knudsen2021experimental},  2021 & MPC & 1 house in Norway  & Radiator & Indoor temperature set-point & 14 days & HVAC energy cost & 22\% vs. simulation \\
\hline

\cite{bunning2022physics},  2022 & MPC & 2 bedrooms in Switzerland & Radiant panels & Water flow & 156 days & HVAC energy & 26--49\% vs. measurement \\
\hline

\cite{vivian2022experimental},  2022 & MPC & 1 house in Italy & Air-to-water heat pump &  Compressor and pump speeds  & 5 days & HVAC energy cost & 10--17\% vs measurement \\
\hline

\cite{wang2023field},  2023 & MPC & 2 rooms in China & Mini-split air conditioners & On-off state & 12 days & HVAC energy cost & 22--27\% vs. measurement \\
\hline

\cite{thorsteinsson2023long},  2023 & MPC & 1  house in Denmark & Air-to-water heat pump & Valve position, outdoor temperature override & 97 days & HVAC energy, emission cost & 2--17\% vs. measurement \\
\hline

\cite{brown2023long},  2023 & MPC & 1 solar house in Canada & Radiant floors & Pump on/off states & 182 days & Temperature regulation  & 71\% vs. measurement \\

\end{tabular}
\end{table*}

\subsection{\color{black} Contributions of this paper} 

{\color{black} This paper makes the following contributions to the literature on supervisory control of residential HVAC systems.

\begin{enumerate}

\item This paper provides a new long-duration field demonstration, the fifth ever to control a real home's HVAC for more than one month.

\item This paper considers an equipment configuration that is common in North America but understudied in the research literature. Specifically, this paper considers a central air-to-air heat pump (4.5 kW rated electric power capacity) with 19 kW of backup electric resistance heating elements that turn on in stages when the heat pump cannot keep up with heat demand. While three residential field demonstrations in the literature considered air-to-air cooling equipment \cite{dong2014real, kurte2020evaluating, wang2023field}, none studied air-to-air heating or resistance backup heat. These missing elements constitute an important research gap, since (a) installers in North America often pair heat pumps with resistance backup heat, especially in cold climates; and (b) this equipment configuration can lead to high peaks in electricity demand in cold weather, when heat demand rises, heat pump efficiencies may fall, and inefficient resistance backup heat may turn on. {\color{black} To handle this equipment configuration within a convex optimization framework, this paper introduces a convex reformulation of the mapping from the heating plant's combined thermal power output to its electrical power input.}

\item {\color{black} This paper develops and field-tests a new, adaptive method for dynamically balancing the competing objectives of energy efficiency and thermal comfort. Using predictions of user preferences and weather conditions, this method periodically adjusts a weight on a thermal discomfort term in the predictive controller's objective function. Field tests, which included regular surveys of real occupants' thermal comfort, showed that this method kept occupants comfortable while yielding substantial energy cost savings.}

\item This paper presents the first residential field demonstration to include reducing peaks in electricity demand as a control objective or to significantly reduce demand peaks. This field evidence builds confidence that advanced HVAC control can mitigate capacity expansion requirements for electrical infrastructure.

\item This paper presents a comprehensive literature review of all published field demonstrations of supervisory control of residential HVAC, identifying several knowledge gaps. 

\item This paper discusses implementation difficulties with supervisory HVAC control, strategies for overcoming those difficulties, and ways to improve scalability by mitigating hardware and labor requirements. While many studies discuss the {\it benefits} of supervisory HVAC control, very few discuss the {\it costs,} which strongly influence economic viability. This paper provides a real-world data point on deployment cost that can inform future benefit/cost analyses.

\end{enumerate}
}

\subsection{\color{black} Organization of this paper} 

Section \ref{reviewSection} of this paper reviews prior field demonstrations of supervisory control for residential HVAC equipment. Section \ref{systemSection} discusses the building, hardware, and software used in our field demonstration. Section \ref{methodsSection} develops the modeling, learning, and control methods underlying the supervisory control system. Section \ref{resultsSection} presents field demonstration results. Section \ref{discussionSection} discusses practical implementation challenges and possible ways to improve scalability. Appendix A defines acronyms and mathematical notation. Appendix B discusses other possible formulations of the optimal control problem considered in \S\ref{methodsSection}.

\section{Review of residential field demonstrations}
\label{reviewSection}

Table \ref{residentialExperimentTable} summarizes 12 field demonstrations of supervisory residential HVAC control. To the best of the authors' knowledge, Table \ref{residentialExperimentTable} includes all residential field demonstration studies published to date. {\color{black} Seven of the 12 studies in Table \ref{residentialExperimentTable} took place in Europe, two in the USA, two in Canada, and one in China. Ten of the 12 studies controlled detached single-family homes, while \cite{bunning2022physics} and \cite{wang2023field} each controlled two rooms within larger buildings. Nine of the 12 studies controlled hydronic heating systems, while \cite{dong2014real} and \cite{kurte2020evaluating} controlled central air conditioners and \cite{wang2023field} controlled mini-split air conditioners. Control actions varied across studies, with some directly manipulating equipment speeds or on/off states and others manipulating set-points of temperatures or fluid flows. Experiment durations ranged from three to 182 days. Eleven of the 12 studies optimized for energy or energy cost; \cite{thorsteinsson2023long} also optimized for greenhouse gas emissions, while \cite{brown2023long} focused on improving temperature regulation in a complex HVAC system for a solar home. Savings in energy or energy cost varied from 2--49\%, with wide variation in performance evaluation practices. This section provides more detail on methods, performance evaluation, and research gaps.}

\subsection{Methods}

\subsubsection{\color{black} Thermal modeling} 

Of the studies listed in Table \ref{residentialExperimentTable}, all but one \cite{kurte2020evaluating} used MPC. All studies modeled thermal dynamics using low-order linear thermal circuit models. This includes \cite{kurte2020evaluating}, which used a thermal circuit model to train an RLC policy offline. The majority of studies used models with 1--3 thermal capacitances and 1--4 resistances. These studies identified model parameters using linear regression, genetic algorithms, particle swarm optimization, or subspace methods for linear state-space models. The amount of training data varied from five days \cite{wang2023field} to 370 days \cite{bunning2022physics}, with most studies using about one month of data. These trends broadly corroborate the findings in \cite{blum2019practical} that low-order thermal circuits trained on 3--5 weeks of data typically perform well enough for control.

\subsubsection{\color{black} Disturbance prediction} 

Disturbances are uncertain, exogenous inputs that influence a system's evolution. Accurate disturbance predictions can significantly improve control performance. The studies in Table \ref{residentialExperimentTable} obtained weather predictions from weather services or forecast models trained on on-site measurements \cite{dong2014real, finck2020optimal}. Some studies predicted solar heat gains using physics-inspired methods that consider solar angles and optical properties of materials \cite{thorsteinsson2023long, brown2023long}. Other studies predicted solar photovoltaic electricity generation \cite{vivian2022experimental} or wholesale electricity market prices \cite{finck2020optimal,thorsteinsson2023long}. Only one study \cite{dong2014real} predicted occupancy.

\subsubsection{\color{black} Sensing} 

Unlike commercial buildings, which often have extensive monitoring and control systems, residential buildings typically have limited sensing infrastructure. With the exception of \cite{wang2023field}, which used Internet-connected outlets with built-in electric power sensors, the studies reviewed in Table \ref{residentialExperimentTable} all measured the thermal power\footnote{This paper uses `thermal power' as a shorthand for the rate of heat transfer to or from a thermodynamic system.} supplied by HVAC equipment. Thermal power sensing typically involves both low-cost temperature sensors and more expensive air or water flow sensors. Two studies \cite{dong2014real, finck2020optimal} measured local weather conditions.

\subsubsection{\color{black} Control actions} 

In the system architecture illustrated in Fig. \ref{architectureFig}, the control action is the signal that the supervisory control system sends to the device-level control system. In the studies in Table \ref{residentialExperimentTable}, most supervisory control systems adjusted the temperature set-point of the indoor air or the water in a storage tank. Some studies adjusted lower-level variables, such as the position of a water flow valve, the on/off state of a heat pump or air conditioner, or the speed of a compressor or pump. In one study, the supervisory control system adjusted a heat pump's perceived outdoor temperature by manipulating the voltage signal from a sensor \cite{thorsteinsson2023long}.

\subsubsection{\color{black} Control objectives} 

All but one of the studies in Table \ref{residentialExperimentTable} aimed to minimize the energy used by HVAC equipment or the cost of HVAC energy. All studies aimed to promote thermal comfort by maintaining the indoor temperature within a given band \cite{dong2014real, afram2017supervisory, brown2023long} or by penalizing deviations of the indoor temperature from the occupants' preferred temperature \cite{thorsteinsson2023long}. Other objectives included maximizing on-site use of solar photovoltaic electricity \cite{vivian2022experimental} and minimizing greenhouse gas emissions under a time-varying greenhouse gas intensity of electricity \cite{thorsteinsson2023long}.

{\color{black}
\subsubsection{Thermal comfort}

Most studies in Table \ref{residentialExperimentTable} sought to maintain occupants' thermal comfort by constraining indoor air temperatures within a band around an occupant-specified preference. No study analyzed thermal comfort using more detailed models such as Predicted Percentage Dissatisfied (PPD)  \cite{van2008forty,enescu2017review}. (PPD models the expected percentage of people who would express discomfort under a given set of conditions, such as air temperature, air flow, clothing level, and activity level.) Only five of the twelve studies controlled spaces with more than a few hours per day of occupancy \cite{pedersen2013central, bunning2020experimental, finck2020optimal, bunning2022physics, thorsteinsson2023long}.  Of these, only \cite{thorsteinsson2023long} reported on occupants' comfort during testing. In \cite{thorsteinsson2023long}, wider temperature constraint bands led to higher energy savings.
}


\subsection{Performance evaluation and savings estimates}

The 12 studies in Table \ref{residentialExperimentTable} reported a wide range of energy and cost savings, from 2--50\%. Some of this variation follows from differences in control algorithms, HVAC equipment, building envelopes, and climates. However, experiment setups and reporting practices also vary in several important ways.

\subsubsection{\color{black} Long vs. short experiments} 

Only five of the 12 studies lasted more than two weeks \cite{pedersen2013central, afram2017supervisory, bunning2022physics, thorsteinsson2023long, brown2023long}. Four of the 12 studies lasted less than one week. Longer experiments yield larger sample sizes and less uncertainty in savings estimates.

\subsubsection{\color{black} Comparison to measurement vs. simulations} 

Eight of the 12 studies compared the measured performance of an advanced supervisory controller to the {\it measured} performance of a benchmark controller \cite{pedersen2013central, bunning2020experimental, finck2020optimal, bunning2022physics, vivian2022experimental, wang2023field, thorsteinsson2023long, brown2023long}. In our experience, this approach typically introduces less uncertainty than comparing to the {\it simulated} performance of a benchmark controller. Even the highest-performing building models often have energy prediction errors of 15\% or more -- model `noise' of similar scale to the savings `signal.' Comparing the measured performance of two controllers brings its own challenges, as weather and occupant behavior vary from day to day. Normalizing performance metrics by weather statistics, such as heating or cooling degree-days, facilitates comparisons across disparate weather conditions.

\subsubsection{\color{black} Whole buildings vs. rooms} 

Three of the 12 studies controlled individual rooms within larger buildings \cite{bunning2020experimental, bunning2022physics, wang2023field}. This approach introduces an accounting risk. In winter, for example, lowering the air temperature in one room will cause heat to flow into it from adjacent, warmer rooms. Maintaining comfortable temperatures within the adjacent rooms will therefore require more energy. Reporting only the energy savings in the controlled room, without accounting for increased energy used to heat adjacent rooms, overestimates the total savings.

\begin{table}
\caption{Duration-weighted average energy cost savings} 
\label{savingsTable} 
\centering
\small
\begin{tabular}{ p{2.3cm} p{2.4cm} p{1.5cm}  }
Compared to measurements? & Controlled whole buildings? & Reported savings \\
\hline
Yes & Yes & 15\% \\
No & Yes & 26\% \\
Yes & No & 36\% \\
\end{tabular}
\end{table}

\begin{figure*}
\centering
\includegraphics[height=0.225\textheight]{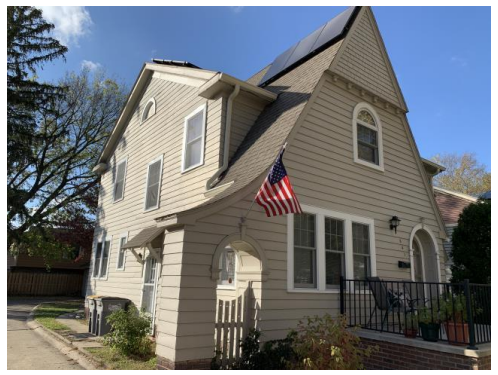} \qquad
\includegraphics[height=0.225\textheight]{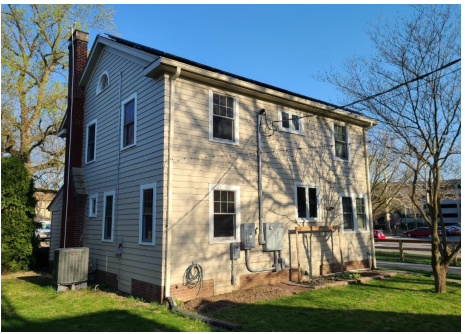}
\label{DCHouseFig}
\caption{The DC Nanogrid House is a 208 m$^2$, 1920s-era house with all-electric appliances in West Lafayette, Indiana, USA.}
\end{figure*}

\subsubsection{\color{black} Current state of reliable savings estimates} 

Of the 12 studies in Table \ref{residentialExperimentTable}, only \cite{pedersen2013central, thorsteinsson2023long, brown2023long} reported experiments that lasted longer than two weeks, compared controller performance to the measured performance of a benchmark controller, and controlled whole buildings. Of those, \cite{pedersen2013central} reported 9\% whole-house energy cost savings over 152 days, \cite{thorsteinsson2023long} reported 2--17\% savings in the combined cost of electricity and greenhouse gas emissions over 97 days, and \cite{brown2023long} did not report savings because the study aimed to improve indoor temperature regulation in a complex HVAC system. Assuming HVAC equipment used about half of whole-house energy, the reported 9\% whole-house energy cost savings in \cite{pedersen2013central} translate to about 18\% HVAC energy cost savings. The two short-duration studies that compared to measurements and controlled whole buildings reported HVAC energy cost savings of 12--15\% over three days \cite{finck2020optimal} and 10--17\% over five days \cite{vivian2022experimental}. The weighted average of the midpoints of the HVAC energy cost savings ranges reported in \cite{pedersen2013central, finck2020optimal, vivian2022experimental, thorsteinsson2023long}, weighted by experiment duration, is 15\%.

\subsubsection{\color{black} Savings overestimation tendencies} 

The duration-weighted average of the midpoint savings estimates in studies that controlled whole buildings, but compared controller performance to simulations rather than measurements, is 26\% \cite{dong2014real, afram2017supervisory, kurte2020evaluating, knudsen2021experimental}. Similarly, the duration-weighted average in studies that compared controller performance to measurements, but controlled individual rooms rather than whole buildings, is 36\% \cite{bunning2020experimental, bunning2022physics, wang2023field}. Relative to the 15\% duration-weighted average in studies that compared performance to measurements and controlled whole buildings, these estimates are higher by factors of 1.7 and 2.4, respectively. Table \ref{savingsTable} summarizes these calculations. While some of the discrepancies in reported savings could trace back to differences in algorithms, prices, buildings, equipment, or climates, the discrepancies are large enough to suggest the possibility of systematic overestimation due to comparing performance to simulations, controlling individual rooms within larger buildings, or both. 

{\color{black} Future work could use statistical methods to formally test the hypotheses that comparing to simulations and/or controlling individual rooms within larger buildings leads to systematic overestimation of savings. If a thorough analysis supported either hypothesis, the research community could propose improved standards for measurement and verification. For example, standards could recommend comparing controller performance only to measurements, not simulations; or comparing to simulations but presenting savings estimates alongside confidence intervals that quantify the accuracy of benchmark models. Standards might also recommend either controlling whole buildings or explicitly accounting for heat transfer between controlled rooms and adjacent thermal zones in savings estimates.}

\subsection{Research gaps}

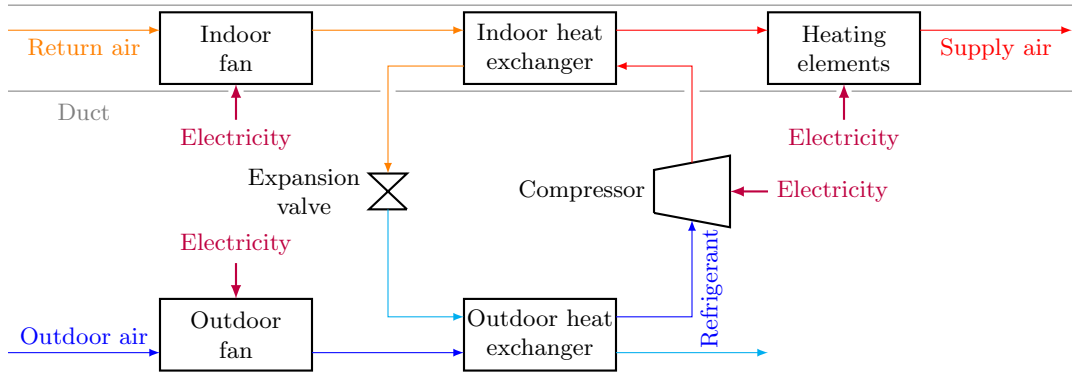
\begin{figure*}
\centering
\small
\begin{tikzpicture}[xscale=1,yscale=0.95]
	\pgfmathsetmacro{\dx}{0.2};
	
	\draw[thick] (-3,4) rectangle (-1,5);
	\node[align=center] at (-2,4.5) {Indoor \\ fan};

	\draw[thick] (1,0) rectangle (3,1);
	\node[align=center] at (2,0.5) {Outdoor heat \\ exchanger};
	
	\draw[thick] (1,4) rectangle (3,5);
	\node[align=center] at (2,4.5) {Indoor heat \\ exchanger};
	
	\draw[thick] (3.5,2+\dx) -- (3.5,3-\dx) -- (4.5,3) -- (4.5,2) -- (3.5,2+\dx);
	\node[align=center,left] at (3.5,2.5) { {\color{white} t} Compressor};
	
	\draw[thick] (0.25,2.25) -- (-0.25,2.25) -- (0.25,2.75) -- (-0.25,2.75) -- (0.25,2.25);
	\node[left,align=center] at (-0.25,2.5) {Expansion\\valve};
	
	\draw[thick] (5,4) rectangle (7,5);
	\node[align=center] at (6,4.5) {Heating \\ elements};
	
	\draw[thick] (-3,0) rectangle (-1,1);
	\node[align=center] at (-2,0.5) {Outdoor \\ fan};

	\draw[blue,->] (3,0.75) -- (4,0.75) -- (4,2.1);
    \node[rotate=90,below,blue] at (4,1.15) {Refrigerant};
	\draw[red,->] (4,2.9) -- (4,4.25) -- (3,4.25);
	\draw[orange,->] (1,4.25) -- (0,4.25) -- (0,2.75);
	\draw[cyan,->] (0,2.25) -- (0,0.75) -- (1,0.75);
	
	\draw[thick,->,purple] (5,2.5) -- (4.5,2.5);
	\node[right,purple] at (5,2.5) {Electricity};
	\draw[thick,->,purple] (6,3.5) -- (6,4);
	\node[below,purple] at (6,3.5) {Electricity};
	\draw[thick,->,purple] (-2,3.5) -- (-2,4);
	\node[below,purple] at (-2,3.5) {Electricity};
	\draw[thick,->,purple] (-2,1.5) -- (-2,1);
	\node[above,purple] at (-2,1.5) {Electricity};
	
	\draw[orange,->] (-5,4.75) -- (-3,4.75);
	\node[below,orange] at (-4,4.75) {Return air};
	\draw[orange,->] (-1,4.75) -- (1,4.75);
	\draw[red,->] (3,4.75) -- (5,4.75);
	\draw[red,->] (7,4.75) -- (9,4.75);
	\node[below,red] at (8,4.75) {Supply air};

    \draw[blue,->] (-5,0.25) -- (-3,0.25);
	\node[above,blue] at (-4,0.25) {Outdoor air};
    \draw[blue,->] (-1,0.25) -- (1,0.25);
    \draw[cyan,->] (3,0.25) -- (5,0.25);

    \draw[gray] (-5,5.1) -- (9,5.1);
    \draw[gray] (-5,3.9) -- (-2.1,3.9);
    \draw[gray] (-1.9,3.9) -- (-0.1,3.9);
    \draw[gray] (0.1,3.9) -- (3.9,3.9);
    \draw[gray] (4.1,3.9) -- (5.9,3.9);
    \draw[gray] (6.1,3.9) -- (9,3.9);
	\node[below,gray] at (-4,3.85) {Duct};

\end{tikzpicture}
\caption{The DC Nanogrid House's HVAC equipment in heating mode. The colors indicate {\color{cyan} cold}, {\color{blue} cool}, {\color{orange} warm}, and {\color{red} hot}.}
\label{HVACFig}
\end{figure*}

{\color{black} Reviewing all available research literature on field demonstrations of supervisory control for residential HVAC equipment identified three knowledge gaps, all of which this paper helps to fill.}

\subsubsection{\color{black} Lack of reliable savings estimates} 

As discussed above, only two field studies \cite{pedersen2013central, thorsteinsson2023long} demonstrated a supervisory residential HVAC controller for more than two weeks, compared its performance to the measured (rather than simulated) performance of a benchmark controller, and either controlled a whole building (rather than rooms within larger buildings) or accounted for increased energy used to condition adjacent rooms. Additional studies that meet these criteria would refine estimates of HVAC energy cost savings. Refined savings estimates could help businesses decide whether to invest in creating supervisory HVAC control products. Refined savings estimates could also help government agencies decide whether and how to incentivize adoption of supervisory HVAC control systems.

\subsubsection{\color{black} Lack of deployment cost and labor estimates}

Deployment costs and labor influence the scalability of supervisory control technology. The residential field demonstration studies in Table \ref{residentialExperimentTable} contain relatively little discussion of deployment costs or labor. Aside from \cite{wang2023field}, which quoted a total hardware cost of 350 Chinese Yuan (about 50 USA dollars), no studies reported how much their sensing, communication, or actuation hardware cost. Similarly, no studies discussed how many hours of engineer labor their field deployments required, or how the labor requirements broke down between tasks. While \cite{blum2022field} and \cite{sturzenegger2015model} discussed deployment costs and labor for commercial buildings, Section \ref{discussionSection} of this paper provides the first thorough discussion in the residential context.

\subsubsection{\color{black} Lack of diversity in objectives and equipment} 

Most studies in Table \ref{residentialExperimentTable} sought to minimize energy costs in a single-device, single-source HVAC system. While energy costs are important, supervisory HVAC control systems can provide other benefits that the field demonstrations in Table \ref{residentialExperimentTable} did not consider. For example, no studies sought to minimize peak electricity demand or maximize flexibility for power grid reliability services. Similarly, no studies considered dual-source HVAC systems, such as electric heat pumps with backup heat from natural gas \cite{beccali2022electrical} or electric resistance. Aside from \cite{vivian2022experimental}, which coordinated a real heat pump with simulated solar photovoltaics, no studies coordinated HVAC systems with other devices. This leaves a gap between field demonstrations and simulation studies that have coordinated HVAC systems with batteries and solar photovoltaics \cite{khakimova2017optimal, langer2020optimal, salpakari2016optimal}, electric vehicles \cite{yousefi2020predictive}, and other devices \cite{gasser2021predictive, sharma2016modeling}. A recent field demonstration in a commercial building showed promise in coordinating HVAC systems with other devices \cite{zhang2022model}.

\subsubsection{\color{black} Treatment of occupant thermal comfort}

{\color{black} Only five studies controlled spaces in residential buildings with more than a few hours per day of occupancy \cite{pedersen2013central, bunning2020experimental, finck2020optimal, bunning2022physics, thorsteinsson2023long}. Of these, only \cite{thorsteinsson2023long} reported on occupants' comfort during testing or investigated the relationship between thermal comfort and energy costs, pollutant emissions, demand peaks, or other objectives. To the authors' knowledge, no study has yet developed methods to adaptively balance trade-offs between thermal comfort and other objectives under dynamic occupant preferences and weather conditions.}

\section{Building, hardware, and software}
\label{systemSection}

\subsection{\color{black} The DC Nanogrid House} 

The field demonstrations in this paper took place in the DC Nanogrid House, pictured in Fig. 2. The DC Nanogrid House is a 208 m$^2$, two-story, 1920s-era detached single-family home near Purdue University's campus in West Lafayette, Indiana, USA. This location falls under the International Energy Conservation Climate Code 5A. This climate sees both hot, humid summers (up to 35 $^\circ$C) and cold winters (down to $-20$ $^\circ$C). We have renovated the DC Nanogrid House with envelope upgrades and all-electric appliances \cite{ore2021evaluation}, including a central air-to-air heat pump with backup resistance heating elements. {\color{black} The exterior walls have foam insulation with an R-Value of 3.5 $^\circ$C m$^2$/W. Code-minimum U-8 W/m$^2$/$^\circ$C windows make up about 20\% of the exterior wall area.} In addition to conventional alternating current wiring, the DC Nanogrid House has a direct current (DC) electrical network that connects its rooftop solar panels, home battery, and DC appliances \cite{ore2021case}. We have installed networked sensors throughout the thermal and electrical systems. The DC Nanogrid House is a living laboratory occupied by graduate engineering students.

\subsection{\color{black} HVAC equipment} 

An air-to-air heat pump, illustrated in Fig. \ref{HVACFig}, provides space conditioning for the DC Nanogrid House. The heat pump has 14 kW of rated cooling capacity, a cooling seasonal coefficient of performance (seasonal COP) of 5.3, and a heating seasonal COP of 2.5. Up to 19.2 kW of backup heating elements turn on in stages (9.6, 14.4, and 19.2 kW) when the heat pump's vapor-compression cycle cannot keep up with heating demand. The heat pump's indoor heat exchanger, the heating elements, and an indoor fan reside within a duct in the basement. The indoor fan blows conditioned air through ducts to 12 vents within the house, drawing air back from six return vents. A second fan blows air over the outdoor heat exchanger. The HVAC system has no outdoor air intake, as infiltration through gaps in the building envelope keeps indoor air fresh. Both fans and the heat pump's compressor have variable-speed drives. To keep the measured indoor temperature near its set-point, the device-level control system (see Fig. \ref{architectureFig}) modulates both fan speeds and the compressor speed roughly in unison. The supervisory control system developed here decides indoor temperature set-points and sends them to the device-level control system.

\subsection{\color{black} Sensing, communication, and computing} 

\begin{figure}
\includegraphics[width=0.45\textwidth]{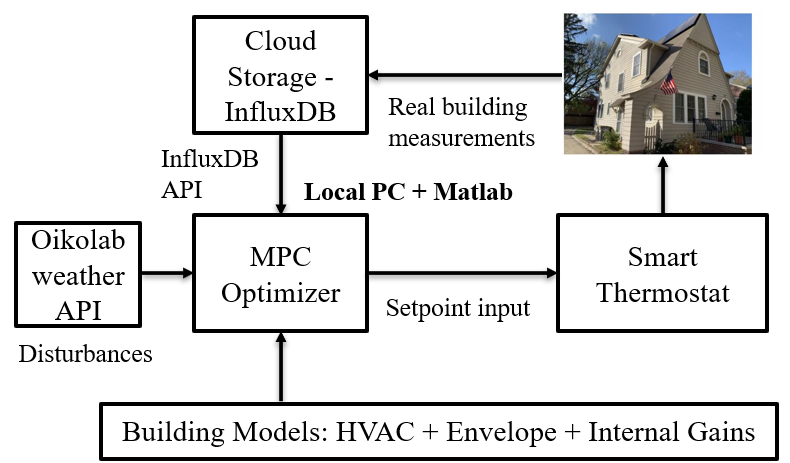}
\caption{Information flow in the control system.}
\label{blockDiagramFigure}
\end{figure}

The supervisory control system uses measurements from Internet-connected power and temperature sensors. An IoTaWatt electric power meter measures a common voltage, as well as individual currents on three circuits that serve the compressor, fans, and heating elements. The electric power meter communicates via Wi-Fi to an InfluxDB database hosted by DigitalOcean. The database also stores the heat pump thermostat's indoor temperature measurement, extracted via Wi-Fi. A desktop computer, located at the DC Nanogrid House, periodically pulls measurements from the database, downloads day-ahead hourly weather forecasts from Oikolab, computes the next indoor temperature set-point in Matlab using the CVX toolbox \cite{cvx}, and pushes the set-point to the thermostat via Wi-Fi and an application programming interface (API). Fig. \ref{blockDiagramFigure} illustrations information flow in the cyber-physical systems.







The sensing, communication, and computing infrastructure described above suffices for supervisory control in practice. For research purposes, we also installed airflow and temperature sensors in the duct upstream of the indoor fan and air temperature sensors downstream of the heating elements (see Fig. \ref{HVACFig}). These sensors provide an estimate of the HVAC plant's thermal power output. The airflow and temperature sensors communicate on serial wires using the Modbus protocol. The wires connect to a Raspberry Pi (a single-board computer with attached input/output circuit pins), which pushes data via Wi-Fi to the InfluxDB database. More information on the sensors and networking can be found in \cite{pergantis2023sensors,ore2020design}.

\section{Modeling, learning, and control methods}
\label{methodsSection}

{\color{black} This section describes the building and HVAC equipment models, the model training methodology, and the MPC formulation. The modeling approach proposed here combines physical intuition with supervised learning techniques for time-series regression. This fusion of physics with machine learning significantly reduces data requirements relative to purely data-driven modeling approaches. In the field, this approach yielded a plant model with sufficient accuracy for control purposes after gathering one month of training data.}

\subsection{Building modeling} 

\subsubsection{\color{black}Thermal circuit model} 

\begin{figure}
\centering
\begin{circuitikz}[scale=1.1, american currents] 
\ctikzset{bipoles/length=1.05cm} 

    \pgfmathsetmacro{\w}{2};
    \pgfmathsetmacro{\h}{1};

    \node[above] at (2*\w,2*\h) {$T_m$};
    \draw (2*\w,2*\h) to[battery1,*-] (2*\w,0) -- (-0.1,0);

    \node[above left] at (\w,2*\h) {$T$};
    \draw (-0.1,2*\h) to (\w,2*\h) to[R,R=$R_m$,*-] (2*\w,2*\h);

    \draw (-0.1,0) to[I,n=Qa] (-0.1, 2*\h);
    \node[right] at (Qa.s) {$\dot Q_c + \dot Q_e$};

    \draw (\w,0) to[C,n=Ca] (\w,2*\h);
    \node[right] at (Ca.s) {$C$};
    \draw (\w,0) node[ground] {} to (\w,0);

    \draw (\w,2*\h) -- (\w,3*\h) to[R,R=$R_\text{out}$,-*] (3*\w,3*\h);
    \draw(3*\w,3*\h) to[battery1] (3*\w,0) -- (2*\w,0);
    \node[above] at (3*\w,3*\h) {$T_\text{out}$};

\end{circuitikz}
\caption{Thermal circuit model of the DC Nanogrid House.}
\label{RCFig}
\end{figure}
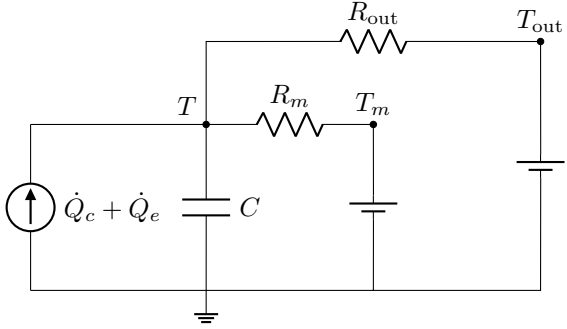

An equivalent thermal circuit model, illustrated in Fig. \ref{RCFig}, captures the DC Nanogrid House's temperature dynamics with sufficient accuracy for control purposes. Thermal circuits behave similarly to electrical circuits, with temperature playing the role of voltage and thermal power playing the role of current. The model in Fig. \ref{RCFig} has one state: the temperature $T$ ($^\circ$C) of the indoor air and shallow thermal mass, such as ducts and the surfaces of walls, floors, and furniture. The model has four input signals: the temperature $T_m$ ($^\circ$C) of the deep thermal mass, such as the interior portions of walls, floors, and furniture; the outdoor air temperature $T_\text{out}$ ($^\circ$C); the thermal power $\dot Q_c$ (kW) from controlled HVAC equipment; and the thermal power $\dot Q_e$ (kW) from exogenous sources such as lights, appliances, electronics, bodies, and the sun. The model has three parameters: the indoor thermal capacitance $C$ (kWh/$^\circ$C), the thermal resistance $R_m$ ($^\circ$C/kW) between the indoor air and deep thermal mass, and the thermal resistance $R_\text{out}$ ($^\circ$C/kW) between the indoor and outdoor air.

\subsubsection{\color{black}Governing equations} 

The indoor temperature dynamics follow the first-order linear ordinary differential equation
\begin{equation}
C \frac{\text d T}{\text d t} = \frac{T_m - T}{R_m} + \frac{T_\text{out} - T}{R_\text{out}} + \dot Q_c + \dot Q_e .
\end{equation}
Defining the effective boundary temperature
\begin{equation}
\theta = \frac{R_\text{out} T_m + R_m T_\text{out} }{R_m + R_\text{out} } \label{effectiveTemperature}
\end{equation}
and the effective thermal resistance
\begin{equation}
R = \frac{R_m R_\text{out}}{ R_m + R_\text{out} } \label{effectiveResistance}
\end{equation}
gives an equivalent differential equation:
\begin{equation}
C \frac{\text d T}{\text d t} = \frac{\theta - T}{R} + \dot Q_c + \dot Q_e . \label{continuousDynamics}
\end{equation}
Assuming the input signals $\theta$, $\dot Q_c$, and $\dot Q_e$ are constant over each time step of uniform duration $\Delta t$ (h), the exact discrete-time dynamics are
\begin{equation}
T(k+1) = a T(k) + (1-a)[ \theta(k) + R(\dot Q_c(k) + \dot Q_e(k))] , \label{discreteDynamics}
\end{equation}
where the integer $k$ indexes time steps and  {\color{black} 
\[
a = \exp\left( -\frac{ \Delta t }{ RC } \right) .
\]
The discrete-time dynamics \eqref{discreteDynamics} come from analytically solving to the first-order linear ordinary differential equation \eqref{continuousDynamics} that governs the continuous-time dynamics. 
}

\subsection{Building model training}

The supervisory control system measures the indoor temperature $T$, downloads measurements of the outdoor temperature $T_\text{out}$ from a weather service, and either measures the HVAC thermal power $\dot Q_c$ or calculates it as the product of the measured electric power and the heat pump COP. Therefore, training the thermal circuit model requires instantiating the deep thermal mass temperature $T_m$, the thermal resistances $R_\text{out}$ and $R_m$, the discrete-time dynamics parameter $a$ (or thermal capacitance $C$), and the exogenous thermal power $\dot Q_e$. We trained the model on passive observations from November 11 to December 10, 2022. These 30 days gave 720 hourly data points for each variable.

\begin{figure*}
\centering
\includegraphics[width=0.45\textwidth]{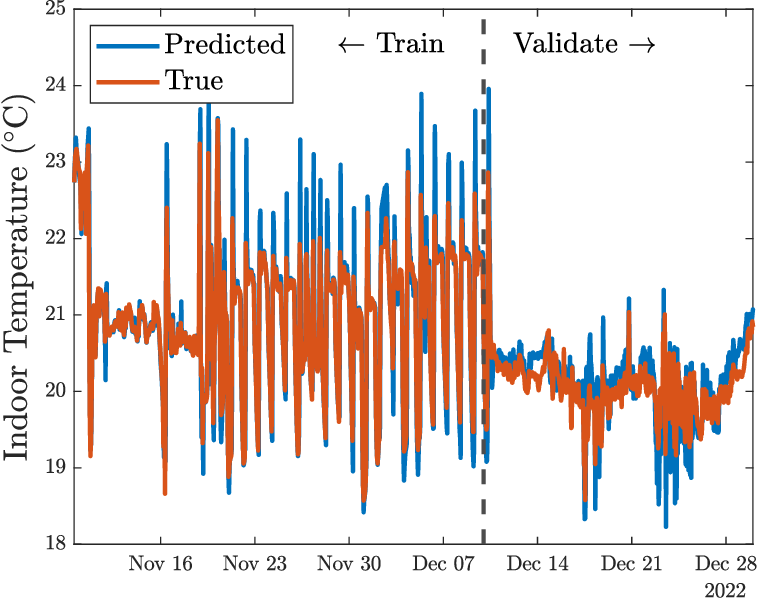}
\qquad
\includegraphics[width=0.45\textwidth]{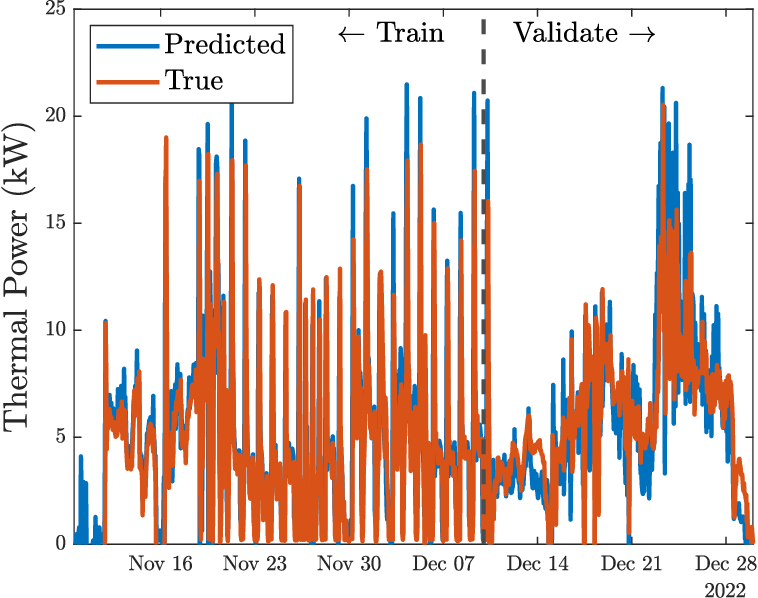}
\caption{The full model's predictions (blue curves) of the indoor temperature (left) and the HVAC thermal power (right) match the targets (orange curves) reasonably well in both training (left two-thirds of data) and validation (right one-third). {\color{black} The validation root-mean-square temperature and power prediction errors were 0.41 $^\circ$C and 2.3 kW, respectively.}}
\label{modelFitFigure}
\end{figure*}

\subsubsection{\color{black} Deep mass temperature} 

Because deep thermal mass temperatures typically change significantly slower than indoor air temperatures, we modeled $T_m$ as constant. {\color{black} This simplification reduced the number of parameters to identify and avoided the need to estimate unmeasured mass temperatures. In buildings with more thermal mass activation, for example through radiant floor heating, the control system could estimate unmeasured thermal mass temperatures using a Kalman filter. However, the constant temperature assumption led to acceptable controller performance with the forced-air heating system in the DC Nanogrid House, consistent with other studies of residential forced-air systems \cite{wang2023field,blum2019practical}. Finally, the modeling approach developed here predicts a catch-all disturbance term that includes unmodeled dynamics such as heat transfer with time-varying thermal mass (see Section \ref{disturbancePrediction} for details).} As most deep thermal mass in the DC Nanogrid House resides in interior material, rather than exterior walls, we estimated the thermal mass temperature by the time-average indoor temperature in the training data, 20.6 $^\circ$C.

\subsubsection{\color{black}Outdoor resistance} 

In steady states with $T_m \approx T$ and approximately constant $\dot Q_e$, the dynamics reduce to
\begin{equation}
T(k) - T_\text{out}(k) \approx \alpha + \dot Q_c(k) R_\text{out} ,
\end{equation}
where $\alpha \approx R_\text{out} \dot Q_e$. To estimate $R_\text{out}$, we formed a dataset from steady times $k_1$, \dots, $k_n$ during nights, when $\dot Q_e$ was approximately constant. The dataset has targets $T(k_i) - T_\text{out}(k_i)$, features $(1, \dot Q_c(k_i))$, and model
\begin{equation}
\bmat
T(k_1) - T_\text{out}(k_1) \\
\vdots \\
T(k_n) - T_\text{out}(k_n) \\
\emat \approx \bmat
1 & \dot Q_c(k_1) \\
\vdots & \vdots \\
1 & \dot Q_c(k_n) \\
\emat \bmat
\alpha \\
R_\text{out} \\
\emat .
\end{equation}
Fitting this model via least squares gave an $R_\text{out}$ estimate of 2.04 $^\circ$C/kW.

\subsubsection{\color{black}Capacitance and mass resistance} 

We fit the capacitance $C$ and resistance $R_m$ using linear regression with a grid search over $R_m$ values from 0.01 to 10 $^\circ$C/kW. Each value of $R_m$ defines an effective boundary temperature $\theta$ and resistance $R$ via Eq. \eqref{effectiveTemperature} and \eqref{effectiveResistance}. Given $\theta$ and $R$, the dynamics during unsteady periods with approximately constant $\dot Q_e$ give
\begin{equation}
\begin{aligned}
&T(k+1) - \theta(k) - R \dot Q_c(k) \\
\approx \ &\beta + ( T(k) - \theta(k) - R \dot Q_c(k) ) a , 
\end{aligned}
\end{equation}
where $\beta \approx (1-a) R \dot Q_e$. As with the outdoor resistance, we fit this linear model via least squares, giving an estimate of $a$ for each value of $R_m$. We split the available unsteady data into training and validation sets. We found that $R_m = 1.06$ $^\circ$C/kW and the corresponding $a = 0.8$ generalized well to the validation data. Although the supervisory control system does not use the capacitance $C = -\Delta t / (R \ln(a))$, the estimate of 6.5 kWh/$^\circ$C, while somewhat higher than expected, matched our physical intuition reasonably well.

\subsubsection{\color{black} Exogenous thermal power prediction} 

\label{disturbancePrediction} After fitting the model parameters, we inverted the dynamics to form a dataset with targets
\begin{equation}
\dot Q_e(k) = \frac{1}{R} \left( \frac{T(k+1) - a T(k)}{1-a} - \theta(k) \right) - \dot Q_c(k) .
\end{equation}
We then followed a standard supervised learning workflow for time-series prediction. We evaluated features including the {\color{black} outdoor temperature, global solar irradiance, wind speed, time of day, day of week}, and past values of the targets.  We compared model structures including linear autoregression, neural networks, regression trees, and support vector machines. We tuned model hyperparameters using two-fold cross-validation. A support vector machine with {\color{black} all of the weather and time features}, but no autoregression, gave an acceptable balance between model complexity and prediction accuracy.

\subsubsection{\color{black}  Trained model performance} 

Fig. \ref{modelFitFigure} shows the end-to-end fit of the full model, including the thermal circuit and the exogenous thermal power predictor, in training and validation data. The one-step-ahead indoor temperature predictions (left plot) match the targets with a root-mean-square error of 0.41 $^\circ$C in the validation data, which we did not use for model training. The HVAC thermal power predictions, formed by substituting the indoor temperature measurements and exogenous thermal power predictions into the dynamics and solving for $\dot Q_c$, matched the measurements with a validation root-mean-square error of 2.3 kW.

\subsection{Heating equipment model}

\begin{figure}
\centering
\includegraphics[width=0.45\textwidth]{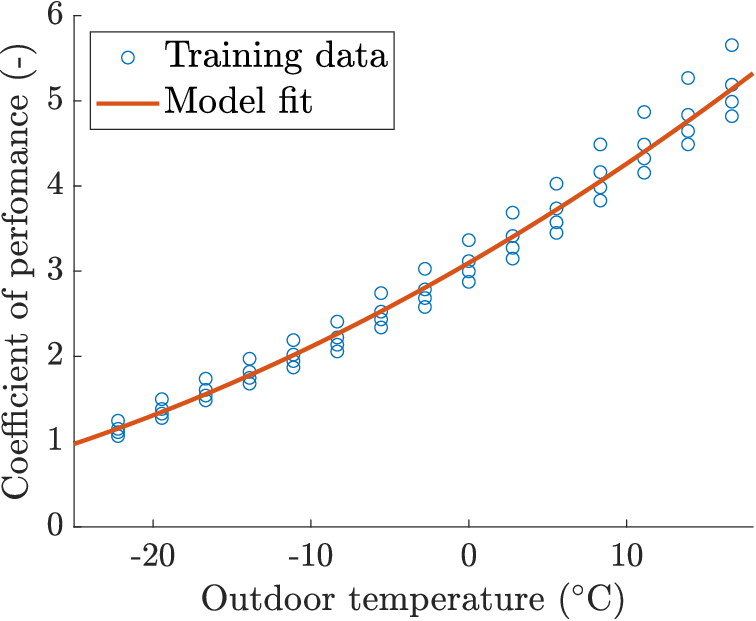}
\caption{The COP model depends only on the outdoor temperature, but matches manufacturer data with $R^2 =$ 0.98.}
\label{COPFig}
\end{figure}

\begin{figure*}
\centering
\begin{tikzpicture}[scale=1]
	\draw[thick,gray] (-0.5,0) -- (3,0);
	\draw[thick,gray] (0,-0.5) -- (0,3);
	\node[below left] at (3,0) {$\dot Q_c$};
	\node[above,rotate=90] at (0,1.85) {Electric power};
	\draw[gray,dashed] (1,0) -- (1,0.33) -- (0,0.33);
	\node[below] at (1,0) {$\eta \overline P$};
	\node[left] at (0,0.33) {$\overline P$};
	\draw[ultra thick] (0,0) -- (1,0.33) -- (3,2.33);
    \draw[dashed] (1.75,1.08) -- (1.75,1.58) -- (2.25,1.58);
    \node[below left] at (1.75,1.58) {1};
	
	\draw[thick,gray] (4.5,0) -- (8,0);
	\draw[thick,gray] (5,-0.5) -- (5,3);
	\node[below left] at (8,0) {$\dot Q_c$};
	\node[above,rotate=90] at (5,1.85) {Electric power};
	\draw[gray,dashed] (6,0) -- (6,0.33) -- (5,0.33);
	\node[below] at (6,0) {$\eta \overline P$};
	\node[left] at (5,0.33) {$\overline P$};
	\draw[ultra thick] (5,0) -- (8,1);
    \draw[dashed] (6.75,0.58) -- (6.75,0.75) -- (7.25,0.75);
    \node[left] at (6.75,0.75) {$1/\eta$};
	
	\draw[thick,gray] (9.5,0) -- (13,0);
	\draw[thick,gray] (10,-0.5) -- (10,3);
	\node[below left] at (13,0) {$\dot Q_c$};
	\node[above,rotate=90] at (10,1.85) {Electric power};
	\node[below] at (11,0) {$\eta \overline P$};
	\draw[ultra thick] (10,0) -- (11,0) -- (13,1.33);
    \draw[dashed] (11.75,0.55) -- (11.75,0.85) -- (12.25,0.85);
    \node[left] at (11.75,0.7) {$1 - 1/\eta$};
	
	\node[] at (1.5,3.5) {$P$};
	\node[] at (3.5,3.5) {$=$};
	\node[] at (6.5,3.5) {$\dot Q_c / \eta$};
	\node[] at (8.5,3.5) {$+$};
	\node[] at (11.5,3.5) {$(1 - 1/\eta) \max \setof{ 0 ,  \dot Q_c - \eta \overline P }$};
	
\end{tikzpicture}
\caption{As a function of the HVAC thermal power $\dot Q_c$, the total electric power $P$ (left) can be decomposed into the sum of a linear function with slope $1/\eta$ (center) and the positive part of a linear function with slope $1 - 1/\eta$ (right). {\color{black} This decomposition allows easy integration with convex optimization software, ensuring efficient global solution of the MPC subproblems.}}
\label{powerFig}
\end{figure*}
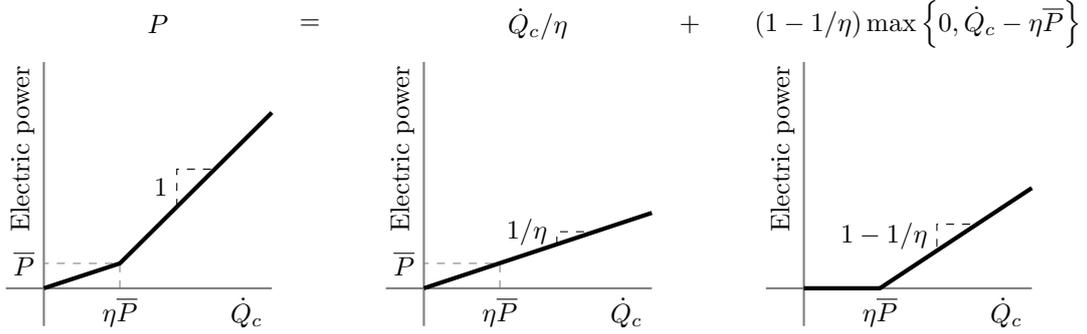

\subsubsection{\color{black}  Heat pump COP} 

In principle, the COP of the DC Nanogrid House's variable-speed air-to-air heat pump depends on the indoor temperature and humidity, outdoor temperature, and compressor speed. In practice, modeling the COP as a function only of the outdoor temperature gave sufficient accuracy for control purposes. Fig. \ref{COPFig} shows how a quadratic COP model fits data from the manufacturer for varying indoor (vertical scatter) and outdoor temperatures. Although the model does not depend on the indoor temperature or compressor speed, it fits the training data with an $R^2$ value of $0.98$.

\subsubsection{\color{black}  Heating plant electric power} 

The heating plant consists of (a) the heat pump with electric power capacity (including the compressor and both fans) $\overline P$ (kW) and COP $\eta$ (a function of the outdoor temperature $T_\text{out}$), and (b) the heating elements with combined electric power capacity $\overline P_r$ (kW) and unity COP. If the heat pump cannot keep up with heating demand, the device-level control system turns on the heating elements. This logic motivates modeling the electric power input to the heating plant as
\begin{equation}
P = \begin{cases}
\dot Q_c / \eta &\text{if } \dot Q_c \leq \eta \overline P \\
\overline P + \dot Q_c - \eta \overline P &\text{otherwise.} \\
\end{cases}
\end{equation}
{\color{black} The above model involves a piecewise definition that does not easily integrate into optimization software without resorting to mixed-integer programming, which significantly increases computational complexity. By contrast,} the equivalent electric power model
\begin{equation}
P = \frac{\dot Q_c}{\eta} + \left( 1 - \frac{1}{\eta} \right) \max \setof{0, \dot Q_c - \eta \overline P}
\end{equation}
integrates directly with convex optimization software. This form shows that $P$ is a convex function of $\dot Q_c$, since the right-hand side is the sum of a linear function and a convex function (the composition of the convex function $\max\setof{0,\cdot}$ with the affine function $\cdot - \eta \overline P$). {\color{black} Integrating directly with convex optimization software guarantees that the MPC subproblems solve in polynomial time to global optimality.} Fig. \ref{powerFig} illustrates the equivalence of the two electric power models.

\subsection{Control algorithm}
\label{controlAlgorithm}

{\color{black} The supervisory control system uses MPC. At each time step, it reads the latest temperature and power measurements, downloads the latest weather forecast from an API service, predicts a trajectory of exogenous thermal powers, solves an open-loop optimal control problem to plan a trajectory of {\color{black} indoor air temperature} set-points, and sends the first planned set-point to the device-level control system. The process repeats at the next time step.}

This subsection describes the open-loop optimal control problem's decision variables, objectives, constraints, and input data. {\color{black} Appendix \ref{otherFormulationsSec} describes several alternative problem formulations that the MPC system could use to pursue other objectives.}

\subsubsection{\color{black}  Decision variables} 

At each time $k$, the supervisory control system solves an open-loop optimal control problem that looks ahead over a receding horizon of $L$ time steps. The $3L$ decision variables in this problem are trajectories over the prediction horizon of the indoor temperature set-points, HVAC thermal powers, and HVAC electric powers:
\begin{equation}
\begin{aligned}
&T(k+1), \dots, T(k+L) \\
&\dot Q_c(k), \dots, \dot Q_c(k+L-1) \\
&P(k), \dots, P(k+L-1) . \\
\end{aligned}
\end{equation}
{\color{black} After solving the optimization problem, the supervisory control system sends the optimal value of $T(k+1)$ to the thermostat as the indoor air temperature set-point for the next time step.}

\subsubsection{\color{black}  Objective function} 

{\color{black} The supervisory control system seeks to minimize the weighted sum of the peak electric power demand cost, the cumulative electrical energy cost, and the cumulative thermal discomfort penalty. The overall objective function is therefore
\begin{equation}
\begin{aligned}
&\pi_d \max \setof{ P(k) , \dots, P(k+L-1) } \\
+ &\Delta t \sum_{\ell = 0}^{L-1} \Big[ \pi_e(k+\ell) P(k+\ell) \\
+ &\pi_t(k+\ell+1) | T(k+\ell+1) - \tilde T(k+\ell+1) | \Big] ,
\end{aligned}
\end{equation}
where $\pi_d$ (\$/kW) is the peak demand price, $\pi_e$ (\$/kWh) is the electrical energy price, $\pi_t$ (\$/$^\circ$C/h) is the thermal discomfort price, and $\tilde T$ ($^\circ$C) is the occupant-specified reference indoor temperature. The control system balances trade-offs between the competing objectives by tuning the peak demand price $\pi_d$ and the thermal discomfort price $\pi_t$. The last paragraph of this section discusses price tuning in more detail.}

\begin{figure*}
\centering
\includegraphics[width=0.45\textwidth]{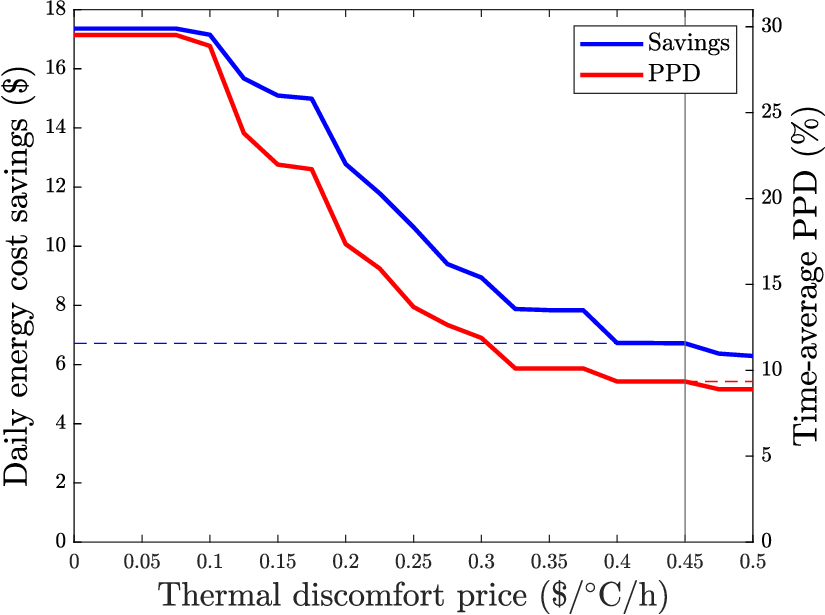}
\quad
\includegraphics[width=0.45\textwidth]{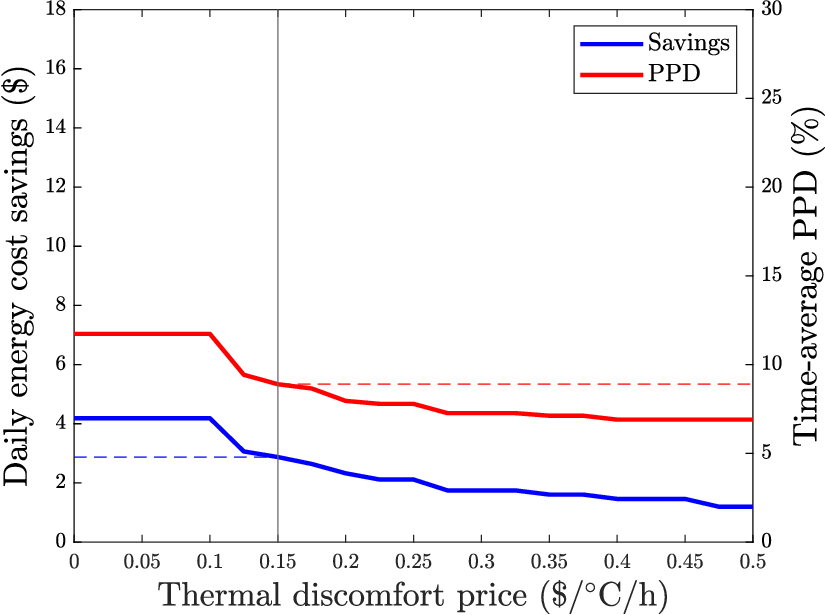}
\caption{The tuned thermal discomfort price (vertical line) was higher on a cold day (left, average outdoor temperature $-8.6$ $^\circ$C) than a mild day (right, 4.8 $^\circ$C). On both days, the simulated time-average PPD was about 10\%.}
\label{discomfortPriceTuning}
\end{figure*}

\subsubsection{\color{black}  Constraints} 

The supervisory control system enforces the following constraints for each $\ell = 0$, \dots, $L-1$: the dynamics
\begin{equation}
\begin{aligned}
&T(k+\ell+1) = a T(k+\ell) + (1-a) [ \theta(k+\ell)  \\
&\quad + R( \dot Q_c(k+\ell) + \dot Q_e(k+\ell) ) ] ,
\end{aligned}
\end{equation}
the heating plant model
\begin{equation}
\begin{aligned}
&P(k+\ell) = \frac{\dot Q_c(k+\ell)}{\eta(k+\ell)} + \left( 1 - \frac{1}{\eta(k+\ell)} \right) \max \Big \{0,  \\
&\quad \dot Q_c(k+\ell) - \eta(k+\ell) \overline P \Big \} , \label{electricPowerDef}
\end{aligned}
\end{equation}
the heating capacity constraints
\begin{equation}
0 \leq \dot Q_c(k+\ell) \leq \eta(k+\ell) \overline P + \overline P_r ,
\end{equation}
and the thermal comfort constraints
\begin{equation}
| T(k+\ell+1) - \tilde T(k+\ell+1) | \leq \delta ,
\end{equation}
where $\delta$ ($^\circ$C) is the maximum permitted absolute deviation of the indoor temperature from the occupant-specified reference temperature $\tilde T$. {\color{black} The time-variation of the COP $\eta(k+\ell)$ in Eq. \ref{electricPowerDef} comes from the dependence of the COP on the outdoor temperature (shown in Fig. \ref{COPFig}). This temperature dependence encourages the optimization to shift thermal load to times of higher COP.}

\subsubsection{\color{black}  Input data} 

In field demonstrations of the supervisory control system, we used time step duration $\Delta t = 1$ h, prediction horizon of $L = 24$ time steps, constant electrical energy price $\pi_e = 0.15$ \$/kWh, heat pump power limit $\overline P = 4.5$ kW, heating element capacity $\overline P_r = 19.2$ kW, temperature deviation $\delta = 3$ $^\circ$C, and trained discrete-time dynamics parameter $a = 0.8$. The trained thermal resistances $R_\text{out} = 2.04$ $^\circ$C/kW and $R_m = 1.06$ $^\circ$C/kW generated the effective thermal resistance $R = 0.7$ $^\circ$C/kW via Eq. \ref{effectiveResistance}, as well as the effective boundary temperature predictions $\theta$ via Eq. \ref{effectiveTemperature} with outdoor temperature predictions $T_\text{out}$ from weather forecasts. We predicted the COPs $\eta$ by propagating the outdoor temperature predictions through the quadratic COP model illustrated in Fig. \ref{COPFig}. We predicted the exogenous thermal power $\dot Q_e$ by propagating weather predictions and time features through the support vector machine described in Section \ref{disturbancePrediction}. The DC Nanogrid House occupants specified reference temperatures $\tilde T$ of 20 $^\circ$C during the day and 18 $^\circ$C overnight.

\subsubsection{\color{black}  Dynamic price tuning} 

We tuned the peak demand price $\pi_d$ and thermal discomfort price $\pi_t$ to balance the competing objectives of energy cost, peak demand, and thermal comfort. {\color{black} The utility serving the DC Nanogrid House does not impose a peak demand charge. Rather, we included the demand price as an additional incentive to mitigate demand peaks by reducing use of backup resistance heat. We fixed $\pi_d$ at \$0.8 per kW of daily peak demand, which corresponds to a monthly peak demand price of about 25 \$/kW, a typical value for monthly demand charges in commercial buildings in the USA. Fixing the peak demand price at a constant value reduced the price tuning problem to calibrating the discomfort price $\pi_t$. Tuning $\pi_t$ interpolates between two extremes: for $\pi_t = 0$, there is no set-point tracking objective; while in the limit $\pi_t \rightarrow \infty$, the set-point is constrained to equal the user's preference $\tilde T$.} {\color{black} The supervisory control system automatically tuned the thermal discomfort price every 12 hours by sweeping an array of $\pi_t$ values, solving an open-loop optimal control problem for each value, selecting the lowest $\pi_t$ that maintained the time-average PPD below 10\%, 
\begin{equation}
    {\color{black}
    \pi_t = \min \{ \pi \mid \text{PPD at price } \pi \leq 10\% \} , }
    \label{price_tuning_equation}
\end{equation}
then multiplying the selected $\pi_t$ by 1.1 during the day and 0.2 overnight. We found that this procedure promoted comfort during daytime hours while allowing greater energy savings overnight. The tuning procedure led to higher discomfort prices on colder days, as Fig. \ref{discomfortPriceTuning} illustrates. 

}

\label{PPDSec}

\section{Field demonstration results}
\label{resultsSection}

\begin{figure}
\includegraphics[width=0.45\textwidth]{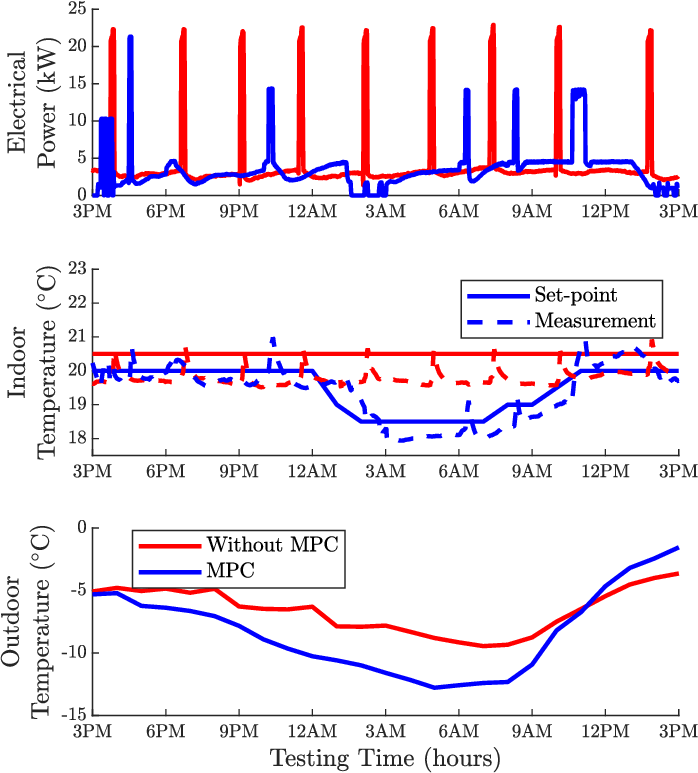}
\caption{The heating equipment used 13\% less energy on a cold day with MPC (blue) than a warmer day with constant temperature set-points (red).}
\label{constantSet-pointComparison}
\end{figure}

{\color{black} This section summarizes field test results from January 30 through March 30, 2023. We tested MPC for 33 of these 59 days. We typically ran MPC for five-day stretches; the longest continuous MPC experiment lasted 12 days. We spent another 15 of the 59 days commissioning and debugging the control system. To benchmark performance, we compared MPC to the remaining 11 of the 59 test days and to historical data from past years. 

We benchmarked MPC performance against the heat pump manufacturer's default device-level control system with occupant-selected indoor temperature set-points and no supervisory control. The manufacturer does not publish the details of their device-level control system for competitive reasons. However, device-level control systems for variable-speed heat pumps typically use Proportional-Integral-Derivative loops that modulate compressor and fan speeds to track an indoor temperature set-point. Manufacturers typically wrap these loops within if-then logic that prevents the compressor from running below $\simm$30\% speed, limits ramping rates, implements minimum on- and off-times, and periodically defrosts the outdoor heat exchanger coils. This logic protects equipment and improves temperature regulation.}

\begin{figure}
\includegraphics[width=0.45\textwidth]{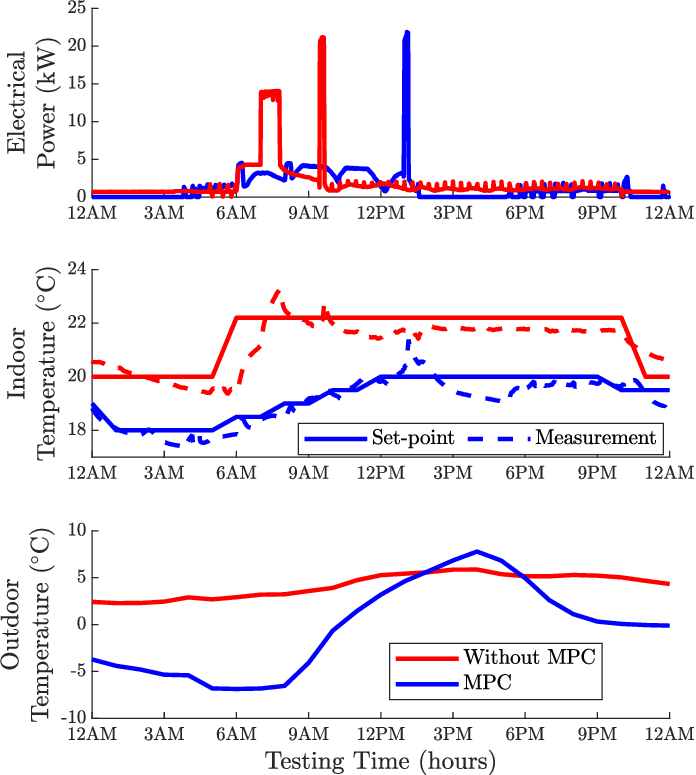}
\caption{The heating equipment used 25\% less energy with MPC on a cold day (blue) than on a warmer day with temperature set-point step changes at 6 AM and 11 PM (red).}
\label{nightSetbackComparison}
\end{figure}

\subsection{\color{black} Control system behavior} 

Fig. \ref{constantSet-pointComparison} compares the control system behavior on a day with MPC (blue curves) to a day with a constant indoor temperature set-point (red curves). On both days, the device-level control system kept the indoor temperature measurements (middle plot, dashed curves) reasonably close to the set-points (solid curves). On the MPC day, the outdoor temperature (bottom plot) reached a low of $-13$ $^\circ$C, compared to $-10$ $^\circ$C on the constant-set-point day. Despite the colder weather on the MPC day, the heating plant used 84 kWh of electrical energy on the MPC day, 13\% less than the 97 kWh used on the constant-set-point day. MPC reduced energy use mainly by reducing the indoor temperature overnight when the outdoor temperature dropped, causing the heat pump COP to drop (see Fig. \ref{COPFig}) and heating demand to rise. MPC reduced the indoor temperature between midnight and three AM, then gradually warmed the house back up between six and nine AM. In addition to reducing energy, MPC reduced the frequency of using high-stage (19.2 kW) backup resistance heat from nine events on the constant-set-point day to one event on the MPC day. However, MPC increased the use of middle-stage (10 kW) resistance heat, primarily during the morning warm-up period.

It may appear from Fig. \ref{constantSet-pointComparison} that simply reducing the indoor temperature set-point overnight could yield similar energy savings to MPC. However, Fig. \ref{nightSetbackComparison} shows that this is not the case in general. Fig. \ref{nightSetbackComparison} compares an MPC experiment day (blue curves) to a non-MPC day (red curves) with two indoor temperature set-points: 22 $^\circ$C during the day and 20 $^\circ$C overnight. Although the MPC day was significantly colder (low of $-7$ $^\circ$C) than the non-MPC day (2 $^\circ$C), MPC used 27\% less HVAC electrical energy than the 44 kWh used without MPC. The heating plant used more energy on the non-MPC day because the backup heating elements turned on in response to the sudden 2 $^\circ$C set-point increase at six AM. While MPC also reduced set-points overnight, it warmed the house up more gradually in the morning, avoiding most use of the heating elements.

\begin{figure}
\includegraphics[width=0.45\textwidth]{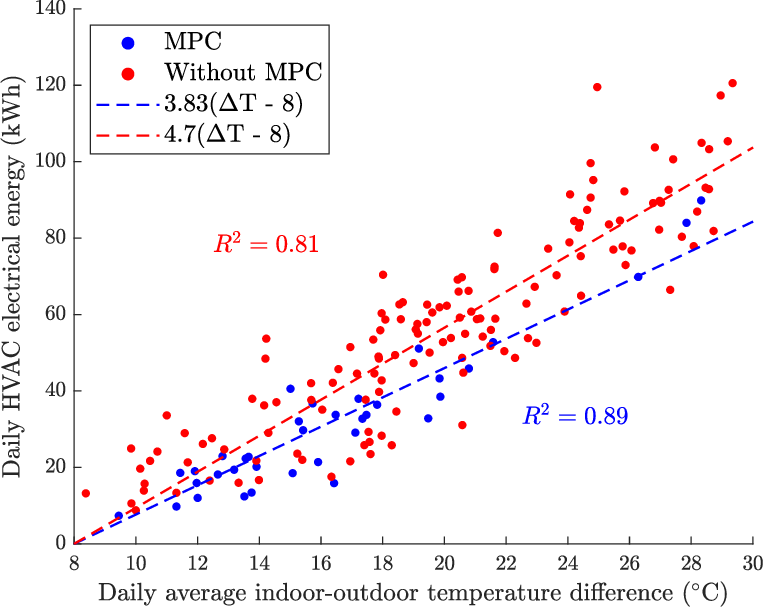}
\caption{MPC typically saves about 20\% of daily HVAC electrical energy. Absolute savings increase with the difference $\Delta T$ between the indoor and outdoor air temperatures.}
\label{energyTemperature}
\end{figure}

\subsection{\color{black} Energy}
\label{energySection}

\begin{figure}
\includegraphics[width=0.475\textwidth]{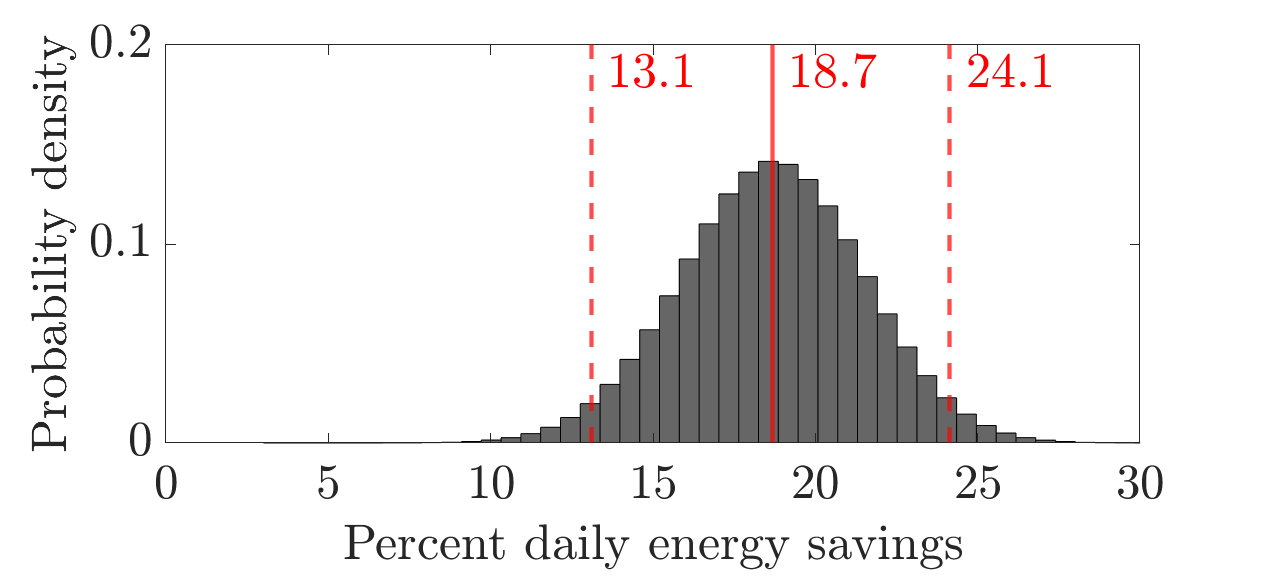}
\caption{Histogram of the percent daily energy savings. Solid red line: mean. Dashed: 95\% confidence interval.}
\label{savingsHistogram}
\end{figure}

Fig. \ref{energyTemperature} shows the daily electrical HVAC energy with MPC (blue) and without MPC (red), as a function of the daily average temperature difference between the indoor and outdoor air. Energy use increases approximately linearly with the temperature difference both with and without MPC, but the slope is higher without MPC. Under the linear fits (dashed lines) in Fig. \ref{energyTemperature}, the slopes are approximately Gaussian distributed. With MPC, the mean and standard deviation of the slope $m$ are 3.83 and 0.117 kWh/$^\circ$C. Without MPC, the mean and standard deviation of the slope $\tilde m$ are 4.71 and 0.076 kWh/$^\circ$C. The relative daily energy savings, as a fraction of the non-MPC daily energy use, are
\begin{equation}
\begin{aligned}
1 - \frac{\text{MPC energy}}{\text{Non-MPC energy}} &\approx 1 - \frac{m (\Delta T - 8) }{\tilde m (\Delta T - 8) }\\
&= 1 - m / \tilde m .
\end{aligned}
\end{equation}
Fig. \ref{savingsHistogram} shows a histogram of the relative savings, based on $10^7$ samples each from the distributions of $m$ and $\tilde m$. The mean relative savings (solid red line) are 18.7\%, with a 95\% confidence interval (dashed red lines) of 13.1 to 24.1\%.

Reduced use of backup heat contributed substantially to the total heating energy savings. On average over 19 days (4 with MPC, 15 without) with daily-average indoor-outdoor temperature differences near 22.5 $^\circ$C, MPC reduced heating element energy use by 38\%, from 15.7 kWh per day to 9.7. On average over these days, backup heat ran for 46 minutes per day with MPC and 55 without. During operation, backup heat tended to run at lower power with MPC than without.

Three effects can explain the observed reduction in backup heat use. First, MPC reduced heating loads by lowering indoor temperature set-points. This let the heat pump more often meet load without backup. Second, MPC sometimes preheated in anticipation of high loads that otherwise would have required backup heat. Third, lower loads during during defrost cycles (discussed further at the end of this section) caused the thermostat to request lower stages of backup heat.

The energy savings estimates presented here are somewhat conservative. A supervisory heating control system can save energy by reducing heating demand, operating equipment more efficiently, or both. Reducing heating demand entails reducing indoor temperatures. Our estimates do not account for energy saved by reducing heating demand as we normalized HVAC energy use by the indoor-outdoor temperature difference.

\subsection{\color{black} Thermal comfort} 
\label{comfort_results}
We assessed the supervisory control system's thermal comfort performance in two ways. First, we periodically surveyed the two occupants using a digital survey. One or both occupants reported feeling uncomfortable on 4.8\% of the 793 test hours with MPC, compared to 4.6\% of the 724 test hours without MPC. Second, we used a thermal comfort software package \cite{tartarini2020pythermalcomfort} to calculate a PPD time series (see Section \ref{PPDSec}) based on indoor temperature measurements and the other inputs to the PPD model, such as occupants' estimated clothing and activity levels. This process yielded a time-average PPD of 10.5\% over the MPC test hours and 9.8\% over the test hours without MPC. (Practitioners typically consider 10\% PPD acceptable \cite{enescu2017review}.) As the occupant comfort reports and PPD calculations produced similar metrics with and without MPC, we concluded that MPC maintained similar thermal comfort to the benchmark control system.

\subsection{\color{black} Demand peaks} 

In the DC Nanogrid House, peaks in electricity used for heating come primarily from the backup heating elements. Some use of heating elements is inevitable because the heat pump is not sized to meet heating demand without backup in the coldest weather. Furthermore, the device-level control system periodically reverses the flow of refrigerant (see Fig. \ref{HVACFig}), drawing heat from the indoor air to melt frost build-up on the outdoor heat exchanger \cite{song2018review}. The device-level control system turns on the heating elements during defrost cycles to avoid blowing cold air on occupants.

\begin{figure}
\includegraphics[width=0.45\textwidth]{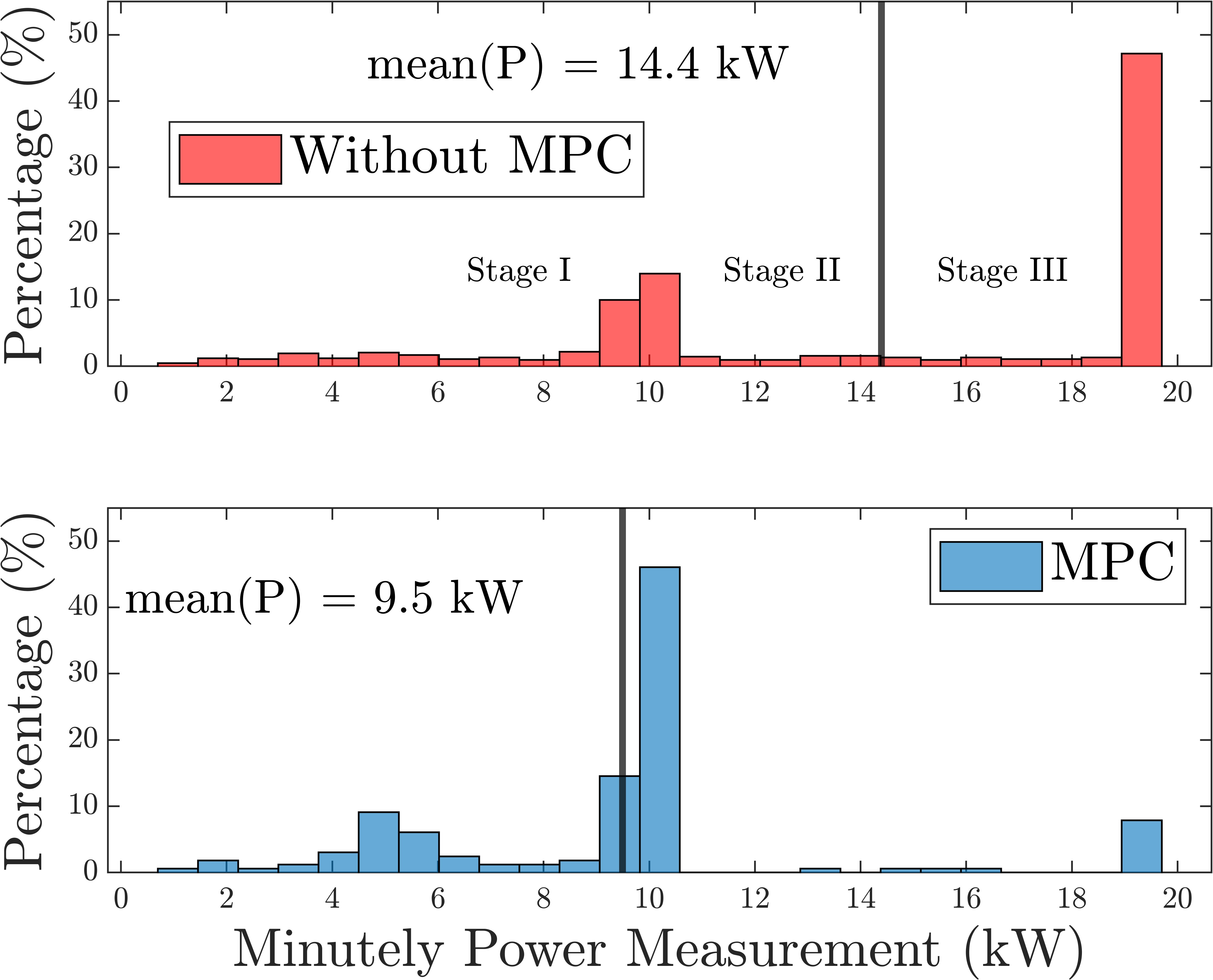}
\caption{Histograms of the heating element power without MPC (top) and with MPC (bottom), conditioned on the event that the heating elements turned on.}
\label{resistanceHistograms}
\end{figure}

While no supervisory control system can completely avoid using the heating elements, MPC reduced the frequency and magnitude of heating element use. In the top plot in Fig. \ref{constantSet-pointComparison}, for example, the frequency of heating element use dropped from nine events per day without MPC (red curve) to six with MPC (blue curve). In the six events where the heating elements turned on under MPC, they turned on at the 9.6 kW stage once, the 14.4 kW stage four times, and the 19.2 kW stage once. Without MPC, the heating elements turned on at 19.2 kW in all nine events. The electric power histograms in Fig. \ref{resistanceHistograms} show that these trends are not unique to the example days in Fig. \ref{constantSet-pointComparison}. Across all field demonstrations, the heating elements ran at 19.2 kW during 48\% of turn-on events without MPC and 8\% with MPC (an 83\% relative reduction). MPC reduced the mean heating element power, conditioned on the event that the heating elements turned on, from 14.7 kW to 9.5 kW.

\subsection{\color{black} Defrost cycles} 

The device-level control system appeared to run fewer defrost cycles with MPC than without. On average over four MPC days with outdoor temperatures below zero $^\circ$C and relative humidities between 70 and 80\% {\color{black} (conditions under which frost often forms)}, the device-level control system ran 1.3 defrost cycles per day. By contrast, the device-level control system ran 2.7 defrost cycles per day over seven comparable days without MPC. {\color{black} We hypothesize that the reduction in defrost cycling came from MPC reducing heat demand in cold, humid weather. However, the sample sizes here are too small to draw robust conclusions. Future research could explore defrost effects in more detail.}

\subsection{\color{black} Seasonal cost savings} 

{\color{black} The temperature-normalized energy models in Fig. \ref{energyTemperature} enable estimation of the total electricity cost savings from MPC in the DC Nanogrid House over the heating season of November 1, 2022, to March 31, 2023. The estimation procedure involves the daily average outdoor temperatures $\bar \theta_i$ ($^\circ$C) over each day $i = 1$, \dots, $151$ of the heating season, the indoor temperature set-point $\tilde T$ ($^\circ$C) without MPC, the difference $\gamma$ ($^\circ$C) between the daily average indoor temperature with and without MPC, the slopes $m$ and $\tilde m$ (kWh/$^\circ$C) from Fig. \ref{energyTemperature}, and the electricity price $\pi_e = 0.15$ \$/kWh. Fig. \ref{energyTemperature} suggests modeling the total electrical energy used for heating on day $i$ without MPC,
\begin{equation}
\tilde E_i = \tilde m \max \{ 0, \tilde T - \bar \theta_i - 8 \} ,
\end{equation}
as zero if the daily average indoor-outdoor temperature difference $\tilde T - \theta_i$ is less than 8 $^\circ$C and linear otherwise. The total electrical energy used for heating on day $i$ with MPC follows a similar model:
\begin{equation}
E_i = m \max \{ 0, \tilde T - \gamma - \bar \theta_i - 8 \} .
\end{equation}

We estimated the seasonal electricity savings from MPC,
\begin{equation}
\tilde E_1 - E_1 + \dots + \tilde E_{151} - E_{151} ,
\end{equation}
using Monte Carlo simulation. The slopes $m$ and $\tilde m$ are normally distributed, as discussed in \S\ref{energySection}. Based on observations over the test days, we set $\tilde T = 20.7$ $^\circ$C and modeled the daily average temperature reduction $\gamma$ between default control and MPC as normally distributed with a 99\% confidence interval of $0.7$ to $1.7$ $^\circ$C. Over 10$^6$ Monte Carlo runs, these parameters gave a mean winter cost savings estimate of \$291 (95\% confidence interval: \$239 to \$363). Relative to the non-MPC winter cost of \$1,029 (\$996 to \$1,061), MPC reduced winter heating costs by 28.3\% (22.7 to 33.7\%).
}




\section{Practical challenges and scalability}
\label{discussionSection}

Section \ref{resultsSection} showed that MPC reduced energy costs and demand peaks while maintaining occupants' thermal comfort. Optimistically extrapolated to all seasons and all residential buildings in the USA, which collectively spend about \$94 billion per year on space heating and cooling energy \cite{RECS2020}, the {\color{black} 23--34\%} heating energy cost savings shown here could translate to {\color{black} \$22--32} billion per year. While substantial, these potential cost savings alone do not justify deployment at scale. Implementation challenges, and the associated costs and labor requirements, also influence the economic case for deployment.

\subsection{\color{black} Practical problems and solutions} 

Deploying our supervisory control system in the field entailed solving a number of practical problems. Establishing communication between a diverse set of devices took time and persistence. In addition to the electric power sensing that our control system required, we installed various sensors for validation purposes. These included flow and temperature sensors throughout the house's thermal systems, which were nontrivial to install, network, and maintain. We recommend that future deployments rely only on electrical sensing and the thermostat's temperature sensor. Pushing indoor temperature set-points to the variable-speed heat pump manufacturer's proprietary thermostat also proved difficult. For future deployments, we recommend installing a thermostat with an open, user-friendly API. Several such thermostats work with most on/off equipment, although most variable-speed equipment today requires a manufacturer's proprietary thermostat. Finally, we found that occupants occasionally felt uncomfortable regardless of the control scheme. We traced this issue to poorly balanced supply air distribution, which we fixed by adjusting damper positions on supply vents. Fixing this issue required trial and error over multiple in-person visits, a labor-intensive process that would not scale well.

\subsection{\color{black} Hardware and installation costs} 

Our field deployment required sensing the HVAC system's electric power use and controlling its indoor temperature set-point. In principle, HVAC electric power can be estimated from whole-home measurements made by an electric utility's smart power meter \cite{lee2021scalable}. In practice, however, electric utilities rarely offer third-party access to real-time smart meter data. Therefore, replicating our field deployment at scale would likely require installing an electric power sensor for each new deployment. A suitable Internet-connected sensor that monitors power on eight circuits currently retails for \$125. Suitable Internet-connected thermostats currently retail for \$50--300. Most power sensor and thermostat installations involve skilled technicians. While costs of technician labor vary, installing a power sensor and thermostat would likely carry similar costs to the devices themselves. At a rough estimate, therefore, hardware and installation might cost several hundred dollars. A business's actual hardware and installation costs would depend on their wholesale (rather than retail) device costs, and on their internal costs of technician labor. An HVAC manufacturer, for example, could outfit equipment with appropriate sensing and actuation hardware in the factory, rather than in the field. This could reduce hardware and installation costs.

\subsection{\color{black} Deployment labor} 

We spent approximately 190 engineer-days deploying, commissioning, and maintaining our supervisory control system. Of that labor, we spent about 150 engineer-days on tasks that we would not have to repeat if we deployed a similar system in another home. These one-time tasks included setting up a cloud database to store all data (low difficulty), writing code to pull data from sensors (medium), building a web form to survey occupant comfort (low), developing general methods to fit building and equipment models (medium), designing a software architecture for computing control actions (medium), developing a general control algorithm (medium), and establishing communication with the thermostat (high). We spent the remaining 40 engineer-days on tasks that we would have to repeat for each additional home. These repeated tasks included installing electric power sensors (0.5 engineer-days), fitting building and equipment models (20), tuning the control system (10), and maintaining the system despite network failures (10). As most team members had not deployed a supervisory HVAC control system before, we expect that labor requirements would decrease if we repeatedly deployed similar systems. 

\subsection{\color{black} Limitations and future work} 

This work spanned one heating season. Running a supervisory HVAC control system for a full year would provide better estimates of annual cost savings. This work also showed an interesting interaction between the supervisory heat pump control system and device-level defrost controls. While it appeared that MPC reduced the rate of frost formation on the outdoor heat exchanger, our sample sizes were small. Longer and more controlled experiments, possibly building on recent frost modeling \cite{ma2023development}, could shed light on these effects. This work also involved little occupant participation in the supervisory HVAC control system. Future work could involve occupants more directly, for example through software that helps a user find their own balance between indoor temperatures, energy costs, greenhouse gas emissions, and impacts on the power grid.


\section*{CRediT authorship contribution statement}

{\bf Elias Pergantis:} Conceptualization, Methodology, Software, Investigation, Formal Analysis, Data Curation, Visualization, Writing – Original Draft, Writing – Review \& Editing. {\bf Priyadarshan:} Methodology, Software, Data Curation, Writing – Original Draft. {\bf Nadah Al Theeb:} Methodology, Data Curation, Formal Analysis, Writing – Original Draft. {\bf Parveen Dhillon:} Methodology, Hardware, Data Analysis - Review, Writing – Review \& Editing. {\bf Jonathan Ore:} Software, Data Curation, Writing – Review \& Editing. {\bf Davide Zivani:} Conceptualization, Methodology, Formal Analysis, Review \& Editing, Project Administration, Funding Acquisition. {\bf Eckhard Groll:} Conceptualization, Methodology, Review \& Editing, Project Administration, Funding Acquisition. {\bf Kevin Kircher:} Conceptualization, Methodology, Formal Analysis, Writing - Review \& Editing, Visualization, Project Administration, Funding Acquisition.

\section*{Declaration of competing interest}

The authors declare that they have no known competing financial interests or personal relationships that could have appeared to influence the work reported in this paper.

\section*{\color{black} Data availability}


{\color{black} The publicly accessible repository \cite{pergantis2023thermal} contains the weather and energy data used to train and validate the thermal models in this paper. The authors will make further data available upon request.}

\section*{Acknowledgments}

The Center for High Performance Buildings (CHPB) at Purdue University supported this work. The Onassis Foundation also supported E. Pergantis as one of its scholars. The authors would like to thank the occupants of the DC Nanogrid House for their patience and availability during the field demonstrations. Additionally, we extend many thanks to Trane Technologies for their technical support throughout the project.

\section*{Appendix A. Acronyms and notation}
\label{notation}

This paper used seven acronyms: HVAC (heating, ventilation, and air conditioning), MPC (model predictive control), RLC (reinforcement learning control), DC (direct current), API (application programming interface), COP (coefficient of performance), and PPD (predicted percentage dissatisfied). Table \ref{notationTable} summarizes the mathematical notation used in this paper.

\begin{table}[t]
\caption{Mathematical notation} 
\label{notationTable} 
\centering
\small

\begin{tabular}{ l | l }
Symbol (Units) & Meaning \\
\hline

$T$ ($^\circ$C) & Indoor air temperature \\
$T_m$ ($^\circ$C) & Thermal mass temperature \\
$T_\text{out}$ ($^\circ$C) & Outdoor air temperature \\
$\dot Q_c$ (kW) & HVAC thermal power \\
$\dot Q_e$ (kW) & Exogenous thermal power \\
$C$ (kWh/$^\circ$C) & Thermal capacitance \\
$R_m$ ($^\circ$C/kW) & Air-mass resistance \\
$R_\text{out}$ ($^\circ$C/kW) & Indoor-outdoor resistance \\
$\theta$ ($^\circ$C) & Effective boundary temperature \\
$R$ ($^\circ$C/kW) & Effective resistance \\
$t$ (h) & Time \\
$\Delta t$ (h) & Time step duration \\
$a$ (-) & Discrete dynamics parameter \\
$k$, $\ell$ (-) & Discrete time indices \\
$\eta$ (-) & HVAC system COP \\
$P$ (kW) & Heat pump electric power \\
$\overline P$ (kW) & Heat pump electrical capacity \\
$\overline P_r$ (kW) & Heating element capacity \\
$L$ (-) & Prediction horizon \\
$\pi_e$ (\$/kWh) & Electricity price \\
$\pi_d$ (\$/kW) & Peak demand price \\
$\pi_t$ (\$/$^\circ$Ch) & Thermal discomfort price \\
$\tilde T$ ($^\circ$C) & Reference indoor temperature \\
$\delta$ ($^\circ$C) & Maximum temperature deviation \\
$u$ ($^\circ$C) & Indoor air temperature set-point \\
$\tau$ (h) & Set-point tracking time constant \\
$\pi_g$ (\$/kg) & Pollutant price \\
$\mu$ (kg/kWh) & Pollutant intensity \\
$\pi_c$ (\$/kWh) & Curtailment price \\
$\hat P$ (kW) & Demand response baseline \\
$m$ (kWh/$^\circ$C) & MPC energy savings slope \\
$\tilde m$ (kWh/$^\circ$C) & Non-MPC energy savings slope \\
$\bar \theta$ ($^\circ$C) & Mean outdoor temperature \\
$\gamma$ ($^\circ$C) & Mean MPC temperature reduction \\
\end{tabular}
\end{table}

{\color{black} \section*{Appendix B. Other problem formulations} 
\label{otherFormulationsSec}

The problem formulation described in \S\ref{controlAlgorithm} reflects the economic incentives and cyber-physical systems of the DC Nanogrid House. With minor modifications, this formulation could accommodate other settings. This section provides a few examples.

\paragraph{Other devices} The supervisory control system developed here could expand to include other Internet-connected devices, such as electric vehicle chargers, home batteries, water heaters, or solar inverters. Adding one of these devices would require modeling the device and fitting the model parameters. By augmenting the open-loop optimal control problem with the new device's model structure and parameters, the supervisory control system could operate the new device in concert with HVAC equipment to achieve house-level objectives, such as reducing peaks in the combined electricity demand of all devices.

\paragraph{Time-varying electricity prices} Although the DC Nanogrid House sees a constant electricity price, some residential customers face time-varying prices. Time-variation can manifest through Time-of-Use rates with two or more price tiers, hourly prices tied to wholesale market prices \cite{kircher2015model}, or Critical Peak Pricing plans that raise prices when the grid is stressed. Given price predictions, the problem formulation described above can accommodate any time-varying pricing scheme.

\paragraph{Imperfect set-point tracking} The control algorithm in Section \ref{controlAlgorithm} implicitly assumes that the device-level control system accurately tracks indoor temperature set-points. We found this assumption accurate as long as set-points changed slower than $\simm$2 $^\circ$C/h. The problem formulation could accommodate imperfect set-point tracking by specifying a tracking model, such as $\text d T / \text d t = ( u - T ) / \tau$, where $u$ ($^\circ$C) is the indoor temperature set-point, $T$ is the actual indoor temperature, and $\tau$ (h) is the tracking time constant; discretizing the tracking model in time; and appending it to the open-loop optimal control problem as a sequence of equality constraints. This formulation would add a trajectory of set-points as decision variables.

\paragraph{Pollutant emissions} The problem formulation could penalize emissions of greenhouse gases or other pollutants by adding the emission cost
\begin{equation}
\pi_g \Delta t \sum_{\ell = 0}^{L-1} \mu(k+\ell) P(k+\ell)  
\end{equation}
to the objective function. This would require a pollutant price $\pi_g$ (\$/kg) and forecasts of the pollutant intensity of electricity, $\mu$ (kg/kWh).

\paragraph{Grid-service revenues} Some retail demand response programs pay for electricity demand reductions below a baseline when the grid is stressed. The problem formulation could include these incentives by subtracting the demand response revenue
\begin{equation}
\Delta t \sum_{\ell = 0}^{L-1} \pi_c(k+\ell) \max \setof{ 0, \hat P(k+\ell) - P(k+\ell) }
\end{equation}
from the objective function. Here $\pi_c$ (\$/kWh) is the curtailment price and $\hat P$ (kW) is the demand response baseline. Similarly, some wholesale electricity markets allow demand-side participants to provide ancillary services such as regulation or reserve \cite{kircher2019heat}. Reserve markets, for example, typically pay demand-side participants in day-ahead markets for committing to quickly ramp power down in response to generator failures or other contingencies. The problem formulation can accommodate wholesale ancillary service revenues through objective terms similar to those for demand response.

\paragraph{Changing formulations over time} The supervisory control system could readily adapt to new economic incentives or changes to cyber-physical systems. For example, if the occupants switched electricity plans, the control system could update the energy prices, peak demand penalties, grid-service incentives, {\it etc.} The control system could automate these adjustments based on input from occupants in a web dashboard or mobile application. The updated policy could go live immediately, eliminating any need to retrain a learning algorithm.
}

\bibliographystyle{elsarticle-num}
\bibliography{refs.bib}

\end{document}